\magnification1095
\input amstex
\hsize32truecc
\vsize44truecc
\input amssym.def
\font\ninerm=plr9 at9truept
\font\twbf=cmbx12
\font\twrm=cmr12

\footline={\hss\ninerm\folio\hss}

\def\section#1{\goodbreak \vskip20pt plus5pt \noindent {\bf #1}\vglue4pt}
\def\eq#1 {\eqno(\text{\rm#1})}
\def\rf#1 {\text{\rm(#1)}}
\let\al\aligned
\let\eal\endaligned
\let\ealn\eqalignno
\let\dsl\displaylines

\let\o\overline
\let\ul\underline
\let\stpr\overrightarrow
\let\a\alpha
\let\b\beta
\let\d\delta
\let\e\varepsilon
\let\f\varphi
\let\D\Delta
\let\g\gamma
\let\G\varGamma
\let\k\kappa
\let\om\omega
\let\P\varPsi

\let\la\lambda
\let\La\varLambda
\let\s\sigma
\let\z\zeta
\let\pa\partial
\let\t\widetilde
\let\u\tilde
\def\sgn{\operatorname{sgn}}

\def\arctg{\operatorname{arctg}}

\def\uP{\ul{\t P}}
\def\RG{\t R(\t\G)}
\def\wiel{2\mu^3+7\mu^2+5\mu+20}
\def\lnz{\ln\(|\z|+\sqrt{\z^2+1}\)+2\z^2+1}
\def\mpl{m\dr{pl}}

\def\({\left(}\def\){\right)}
\def\[{\left[}\def\]{\right]}
\def\dr#1{_{\text{\rm#1}}}
\def\gi#1{^{\text{\rm#1}}}
\def\ap{approximation}
\def\ce{contemporary epoch}
\def\ct{constant}
\def\co{cosmological}
\def\dc{dependen}
\def\ef{effective}

\def\gr{gravitational}
\def\il{inflation}

\def\pt{parameter}
\def\pc{particle}
\def\q{quint\-essence}
\def\sy{symmetr}
\def\tr{transform}
\def\U{Universe}
\def\wrt{with respect to }
\def\KK{Kaluza--Klein}
\def\JT{Jordan--Thiry}
\def\nos{nonsymmetric}
\def\up#1{\uppercase{#1}}
\def\eu{\expandafter\up}
\def\dint{-\kern-11pt\intop}
\def\tin#1{\hbox to\parindent{\hfil #1)\enspace}}

{\twbf
\advance\baselineskip4pt
\centerline{M. W. Kalinowski}
\centerline{\twrm (Higher Vocational State School in Che\l m, Poland)}
\centerline{Comment on a Quintessence Particle Mass}
\centerline{in the Kaluza--Klein Theory}
\centerline{and Properties of its Field}}

\vskip20pt plus5pt

{\bf Abstract.} We calculate a mass of a \q\ particle of order $10^{-5}\,$eV
and we find several solutions for \q\ field equation. We consider also a \q\
speed of sound in several schemes and \q\ fluctuations.

\vskip20pt plus5pt

In this paper we consider several consequences of the \eu\nos\ \KK\ (\JT)
Theory. We consider a value of a mass of \q\ \pc, several interesting
relations among energy scales, radiation density in the second de Sitter phase.
We find a spatial dependence of a \q\ field (and an \ef\ \gr\ \ct~$G\dr{eff}$)
in a case of spherical-statical \sy y, cylindrical-statical \sy y,
flat-static \sy y. We find a time \dc ces of a \q\ field (with no spatial
\dc ce). We get a solution for a \q\ field (a~travelling wave) and
two-dimensional wave  solution applying those solutions for~$G\dr{eff}$. We
propose some kind of statistical approach to our results. We calculate a
speed of sound in a \q. We consider also fluctuations of a \q\ caused
by \gr\ waves perturbations.

Let us consider a value of mass of a \q\ \pc\ (a~scalar \pc) (see Ref.~[1])
obtained from \eu\nos\ \KK\ (\JT) Theory (see Ref.~[2]):
$$
m_0=\sqrt{\frac n2}\(\frac n{n+2}\)^{n/4} |\g|^{1/2}
\(\frac{|\g|}\b \)^{n/4}. \eq1
$$
(see Eq.~(79) from Ref.~[1]).

The value of this mass has been obtained by this
\pc\ during the second de Sitter phase. Moreover during our contemporary
epoch it is the same. The \pt s $n$, $\g$ and~$\b$ are defined in Refs
[1],~[2]. During the second de Sitter phase a \co\ \ct\ has been calculated
in Ref.~[2] and one finds
$$
\la_{c0}=6H_1^2=\frac{2|\g|^{(n+2)/2}n^{n/2}}{\b^{n/2}(n+2)^{(n+2)/2}}\,.
\eq2
$$
The value of a \co\ \ct\ remains the same during our contemporary epoch.
$H_1$~is a Hubble \ct\ during the second de Sitter phase (see Eq.~(8) of
Ref.~[1]). 

Thus for our contemporary epoch one gets
$$
m_0=\frac12 \sqrt{n(n+2)}\,\sqrt{\la_{c0}}\,. \eq3
$$
In this way we can connect the value of a \co\ \ct\ of our contemporary epoch
to the value of a mass of a \q\ \pc. From recent observational data we get
$$
\la_{c0}=\La=10^{-52}\frac1{\text{\rm m}^2}\,. \eq4
$$
Moreover in order to get a correct dimension for a mass we should add a
factor with~$\hbar$ and~$c$ (now we abandon the system of units with
$\hbar=c=1$).

One gets
$$
m_0=\frac12 \frac{\hbar}c \,\sqrt{n(n+2)}\,\sqrt{\la_{c0}}\,. \eq3a
$$
Using the value of $\la_{c0}$ one finally gets
$$
m_0\simeq \sqrt{n(n+2)}\cdot 0.17\cdot 10^{-39}\,\text{g} \eq5
$$
or
$$
m_0\simeq \sqrt{n(n+2)}\cdot 0.95\cdot 10^{-5}\,\text{eV}. \eq5a
$$
For example, if we take $n=14(=\dim G2)$, one finally gets
$$
m_0\simeq 14.2\cdot 10^{-5}\,\text{eV}. \eq6
$$

This value is bigger than that considered by different authors. Moreover, 
still sufficiently small. The \pc\ interacts only gravitationally and because
of this it is undetectable by using known experimental methods.
Taking a density of dark energy as $0.7$ of a critical density,
$$
\rho_c=1.88 h^2 \cdot 10^{-29}\frac{\text{g}}{\text{cm}^3}\,, \eq7
$$
one gets a number of \q\ \pc s per unit volume
$$
\o n=\frac{h^2}{\sqrt{n(n+2)}}\cdot1.31 \cdot10^{10}\, \frac1{\text{cm}^3} \eq8
$$
where $h$ is a dimensionless Hubble \ct\ $0.7<h<1$. Taking $n=14$ and $h=0.7$
one finally gets
$$
\o n=4\cdot10^8\,\frac1{\text{cm}^3} \eq8a
$$
which is many orders of magnitude smaller than Loschmidt number. Thus a gas
of \q\ \pc s is not so dense from the point of view of our earth conditions.
However, if this number of \pc s per unit volume is considered in a container of
size 200\,Mpc, the gas can be considered as extremely dense.

In order to settle---is this gas dense or not---we should calculate a mean
scattering length. The scattering cross-section for a \q\ \pc
$$
\sigma=\frac1{\la_{c0}}=10^{52}\,\text{m}^2. \eqno(*)
$$
A mean scattering length
$$
l=\frac1{\sigma n} \eqno(*{*})
$$
where $n$ is a number of \q\ \pc s per unit volume (Eq.~\rf8a ).

One gets
$$
l=10^{-60}\,\text{m}. \eqno(*{*}*)
$$
It means that a gas of \q\ \pc s is extremely dense (if we apply the Knudsen
criterion---a gas is dense if $l\ll L$, where $L$ is the size of the
container) even in the Solar System.

Let us consider the Eq.~(111) from Ref.~[3] which connects several scales of
energy and gives an account of the smallness of \gr\ interactions in our
theory. We rewrite this equation in the form
$$
\(\frac{m\dr{pl}}{m\dr{EW}}\)\(\frac{m_{\u A}}{m\dr{EW}}\)^{n_1}=
\(\frac{n|\g|}{(n+2)\b}\)^{(n+2)(n_1+2)/2} 
\eq9
$$
where $m\dr{EW}$ is an electro-weak interactions energy scale. Taking
$$
\al
&m_{\u A}=m\dr{EW}\\
&n_1=2\ (M=S^2)\\
&n=14\ (H=G2)
\eal
$$
one gets
$$
\frac{m\dr{pl}}{m\dr{EW}}=\(\frac78\,\k\)^{24} \eq10
$$
where
$$
\k=\frac{|\g|}{\b} \eq11
$$
and eventually one gets
$$
\k=\(\frac{m\dr{pl}}{m\dr{EW}}\)^{1/24}\cdot\frac87\,. \eq12
$$
Taking $m\dr{EW}\simeq80\,$GeV and $m\dr{pl}\simeq2.4\cdot10^{18}\,$GeV one
gets 
$$
\k\simeq6.19 \eq13
$$
which is very reasonable for it establishes a relation between two \co\ terms
$\g$ and~$\b$ as of the same order. Simultaneously this is a consistency
condition for our model with energy scales (the 6-dimensional Planck's mass
is equal to~$m\dr{EW}$). In this way we can calculate a mass of a \q\ \pc\
for our \ce\ from Eq.~\rf1
$$
m_0=2\cdot10^{15}\,\text{eV}\cdot|\uP|^{1/2}. \eq14
$$
This gives us an estimation for $|\uP|$ and $\RG$:
$$
\RG=\frac1\k \(\frac{m\dr{EW}}{m\dr{pl}\a\dr{em}}\)^2|\uP| \eq15
$$
where $\a\dr{em}=\frac1{137}$ is a fine coupling \ct. Using Eq.~\rf13 \ and
values of $m\dr{pl}$ and $m\dr{EW}$ one gets
$$
\RG=24.75\cdot 10^{-31}|\uP|. \eq16
$$
From Eq.\ \rf14 \ and Eq.\ \rf6 \ one gets
$$
|\uP|=5\cdot10^{-39} \eq17
$$
and
$$
\RG\cong 1.2\cdot 10^{-68}. \eq18
$$
Moreover in our simplified theory we have
$$
\RG=\frac{2(\wiel)}{(\mu^2+4)^2} \eq19
$$
and $\mu$ should be very close to the root of the polynomial
$$
W(\mu)=\wiel. \eq20
$$
From \rf18 \ and \rf19 \ one gets
$$
\wiel\cong1.3\cdot10^{-66}. \eq21
$$
Thus $\mu$ is very close to the 70-digit approximation of the root of the
polynomial \rf20 \ ($W(\o \mu)=0.1\cdot10^{-67}$).
Due to this we can control the \co\ terms.

In the case of $|\uP|$ we have the formula (28) from Ref.~[1] and one gets
$$
\ealn{
&\uP(\z_0+\e)\cong \uP(\z_0)+\frac{d\uP}{d\z}(\z_0)\e &\rf22 \cr
&\uP(\z_0=\pm1.38\ldots)=0 &\rf23 \cr
&\Bigl|\frac{d\uP}{d\z}(|\z_0|=1.38\ldots)\Bigr|=25. &\rf24 
}
$$
Thus one gets from \rf17 \ and \rf23--24 
$$
\e\simeq 2\cdot10^{-40}. \eq25
$$
Thus we need an \ap\ of $\z_0$ up 40-digit arithmetics. We see that \co\
terms coming from the \eu\nos\ \KK\ (\JT) Theory are very small, but not
zero, and that they are easily controllable by $\mu$ and~$\z$ \pt s.

Let us consider a self-interaction potential for a \q\ field for our \ce\
(which is the same as for the second de Sitter phase). One gets from \co\
terms $\la_{c0}(\P)$
$$
\ealn{
\la_{c0}(\P)&=-\frac12\(\b e^{2\P}-|\g|\)e^{n\P} &\rf26 \cr
U(q_0)&=-\frac{|\g|}{2(n+2)}\(\frac{n|\g|}{(n+2)\b}\)^\frac n2
\exp\(\frac{nm\dr{pl}}{2\sqrt{2\pi|\o M|}}q_0\)\cr
&\qquad{}\times\(n\exp\(\frac{m\dr{pl}}{\sqrt{2\pi|\o M|}}q_0\)-(n+2)\) &\rf27 \cr
}
$$
or
$$
U(q_0)=-\frac14\la_{c0}
\exp\(\frac{nm\dr{pl}}{2\sqrt{2\pi|\o M|}}q_0\)
\(n\exp\(\frac{m\dr{pl}}{\sqrt{2\pi|\o M|}}q_0\)-(n+2)\) 
\eq27a
$$
where $\la_{c0}$ is a \co\ \ct\ for our \ce\ and
$$
\P=\P_0+\frac{m\dr{pl}}{2\sqrt{2\pi|\o M|}}q_0=\P_0+\o\b q_0=\P_0+\f. \eq28
$$
For $\la_{c0}$ is very small ($10^{-52}\frac1{\text{m}^2}$), this interaction
is very small. For $n=14(=\dim H=\dim G2)$ one gets
$$
U(q_0)=-\frac12\la_{c0}
\exp\(\frac{7m\dr{pl}}{\sqrt{2\pi|\o M|}}q_0\)
\(7\exp\(\frac{m\dr{pl}}{\sqrt{2\pi|\o M|}}q_0\)-8\) .
\eq29
$$

Even in the potential $U(q_0)$ we have exponential terms, the strength of the
interactions is negligible for small value of~$q_0$. The interesting point in
our theory is the \ef\ \gr\ \ct\ (depending on a scalar field~$\P$). Let us
describe it by a \q\ field~$q_0$. One gets
$$
G\dr{eff}=G_0e^{-(n+2)\P}. \eq30
$$
After some calculations one finds
$$
G\dr{eff}=G_0\(\frac{(n+2)\b}{n|\g|}\)^{n+2}
\exp\(-\frac{n+2}{2} \frac{m\dr{pl}}{\sqrt{2\pi|\o M|}}q_0\) \eq31
$$
or
$$
G\dr{eff}=\frac{G_0}{\la_{c0}^2}\cdot \frac{4(n+2)^2\b^2}{n^2}
\exp\(-\frac{n+2}{2} \frac{m\dr{pl}}{\sqrt{2\pi|\o M|}}q_0\). \eq31a
$$
It is easy to see that the r\^ole of a Newton's \ct~$G_N$ is played by
$$
G_N=\frac{G_0}{\la_{c0}^2}\(\frac{2(n+2)\b}{n}\)^2. \eq32
$$
This is the reason that we put a \ct\ $G_0$ in the formula \rf30 .

One gets
$$
G_0=G_N\la_{c0}^2\(\frac{n}{2(n+2)\b}\)^2. \eq33
$$
For 
$$
\b=\a^2_sm\dr{pl}^2\RG=\a^2_s\frac{\RG}{l\dr{pl}^2} \eq34
$$
one gets
$$
G_N=\frac{G_0\a_s^4}{(\la_{c0}l\dr{pl}^2)^2}\(\frac{2(n+2)\RG}{n}\)^2 \eq35
$$
or
$$
G_0=G_N\(\frac{n}{2(n+2)\RG}\)^2\(\frac{\la_{c0}l\dr{pl}^2}{\a_s^2}\)^2. \eq36
$$

Let us connect to $G_0$ a new Planck's mass $m^0\dr{pl}$. In terms of this
mass and ordinary $m\dr{pl}$ mass one gets
$$
m^0\dr{pl}=m\dr{pl}\,\frac{2(n+2)\RG}{n}\,\frac{\a^2_s}{\la_{c0}l^2\dr{pl}}
\eq37
$$
or
$$
m\dr{pl}=m^0\dr{pl}\,\frac{n}{2(n+2)\RG}\,\frac{\la_{c0}l^2\dr{pl}}{\a^2_s}\,.
\eq38
$$

If we take our simplified model with $\RG$ given by Eq.~\rf19
$$
\a_s^2=\a\dr{em}=\tfrac1{137},\ n=14,
$$
we get
$$
\mpl^0=\mpl\,\frac{32(\wiel)}{7(\mu^2+4)^2}\,\frac{\a\dr{em}}{\la_{c0}
l^2\dr{pl}} \eq39
$$
or
$$
G_0=G_N\(\frac{7(\mu^2+4)^2}{32(\wiel)}\)^2\(\frac{\la_{c0}l^2\dr{pl}}
{\a\dr{em}}\)^2. \eq40
$$
Moreover we can connect unknown \ct~$G_0$ ($\mpl^0$) with an energy scale of
electro-weak interactions $m\dr{EW}$ in a way suggested in Ref.~[3] and
developed here in this paper.

It is enough to use the formula \rf10 \ and we get
$$
\mpl^0=m\dr{EW}\(\frac{7\k}8\)^{24}
\frac{32(\wiel)}{7(\mu^2+4)^2}\,\frac{\a\dr{em}}{\la_{c0}l^2\dr{pl}}. \eq41
$$
If we take for $\la_{c0}$ the known value of the \co\ \ct\ (Eq.~\rf4 ) for
$l\dr{pl}=4\cdot 10^{-32}\,\text{cm}$ and $\k=6.19$ we get
$$
\mpl^0=8.12\cdot10^{132}\cdot\frac{\wiel}{(\mu^2+4)^2}\,m\dr{EW}\,. \eq42
$$
Moreover we see that $\mu$ must be very close to the root of the polynomial
$\wiel$ (see Ref.~[3]) and we get
$$
\mpl^0=2\cdot10^{130}(\wiel)m\dr{EW}\,. \eq43
$$
If we take 70-digit \ap\ of the root of the polynomial \rf20 \ and considered
value of $W(\mu)$ (Eq.~\rf18 ) obtained here, we get
$$
\mpl^0=4.6\cdot 10^{64}m\dr{EW} \eq44
$$
or
$$
\mpl^0=4.6\cdot 10^{64}\,\frac{m\dr{EW}}{\mpl}\,\mpl=
1.5\cdot 10^{48}\mpl = 3.2\cdot 10^{43}\text{g}=1.6\cdot10^{10}M_\odot
\eq45
$$
where $M_\odot=1.99\cdot10^{33}\,$g is a Solar mass. (This is a mass of our
Galaxy.)

Moreover the present critical density of matter in the \U\ is
$$
\rho_{c,0}=2.775h^{-1}\times 10^{11}\,\frac{M_\odot}{(h^{-1}\,\text{Mpc})^3}\,.
\eq46
$$
So we can find a volume containing $\mpl^0$. One gets
$$
V_0=\frac{\mpl^0}{\rho_{c,0}}=5\cdot10^{-3}h^2(\text{Mpc})^3. \eq47
$$
Thus the size of this volume is of order
$$
L_0\simeq \root 3 \of{h^2}\,0.17\,\text{Mpc}. \eq48
$$

Taking under consideration the fact that our calculations are quite rough
(for example $\k$ could be bigger a little, or $W(\t \mu)$ a little bigger),
we come to the conclusion that $\mpl^0$ is of order of the mass of our visible
\U\ (a~volume of size of $10^4$\,Mpc). In this way it seems natural to suppose that
$$
\mpl^0=M_U \eq49
$$
where $M_U$ is the total mass of the visible \U. However, this intriguing
conjecture demands many investigations.

Moreover it is in a real spirit of Dirac's large number hypothesis
and can give a link between cosmology and fundamental interactions theory.
Thus if we suppose \rf49 \ we get
$$
G_0=G_U=\frac1{M_U^2}\,. \eq50
$$
Let us come back to the equation for an \ef\ \gr\ \ct\ (Eq.~\rf32 ):
$$
G\dr{eff}=G_N \exp\(-\frac{n+2}2\,\frac{\mpl}{\sqrt{2\pi |\o M|}}\,q_0\).
\eq51
$$
$q_0$---a \q\ scalar field---possesses a mass and because of this it has a
finite range with Youkave type behaviour
$$
q_0=\frac{\a}{R}\exp\(-\frac R{r_0}\) \eq52
$$
where 
$$
r_0=\frac2{\sqrt{n(n+2)}\sqrt{\la_{c0}}}\quad\hbox{and $\a$ is a positive
\ct} \eq53
$$
or taking for $n=14$ and for $\la_{c0}$ Eq.~\rf4 ,
$$
r_0=3\cdot 10^{23}\,\text{m}=10\,\text{Mpc}. \eq54
$$
The formula \rf52 \ is valid for $R\simeq r_0$. Thus we cannot observe scalar
\gr\ radiation from closed binary sources and the quadrupole radiation
formula is satisfied for a \gr\ field.

Moreover for $R<r_0$ we should take under consideration a full
self-interaction potential of a field~$q_0$ (Eq.~\rf29 ). Moreover, because
of the \ct\ $\la_{c0}$ in front of the formula \rf29 \ this is negligible.
Because of this we can repeat some considerations from the first point of
Ref.~[2] coming back to the field~$\P$ and consider different sources of mass
for this field due to interaction with the matter. In this way in both cases
$$
G\dr{eff}=G_N\,. \eq55
$$
However, we can expect some small effects on short distances.

Let us notice that in the previous \ap\ we consider a weak field, it means
$|q_0|\ll 1$. This is different from small field. A small field considered
below is such that $q_0$ is small in a usual sense. It is negative. Moreover
it can be big in the sense of the absolute value.

Let us calculate a radiation density in the moment when the second de Sitter
phase starts. It means the radiation released after the phase transition. One
gets
$$
\rho_r=\la_{c1}(\P_1)-\la_{c0}(\P_0)=
\la_{c0}(\P_0)\(\frac{H_0^2}{H_1^2}-1\). \eq56
$$
Using Eqs (1a) and (8a) from Ref.\ [2] one gets
$$
\al
\rho_r&=\la_{c0}(\P_0)\cr
&\times \(\frac{n^2|\uP|^2+|\uP|\sqrt{n^2|\uP|^2+4(n^2-4)A\RG}
+4A\RG(n+2)}{2^{(n+2)/2}|\uP|^{(n+2)/2}n^{n/2}}\right.\cr
&\qquad{}\times\left.\vphantom{\frac{\sqrt{|\uP|^2}}{|\uP|^{(n+2)/2}}}
\(n|\uP|+\sqrt{n^2|\uP|^2+4(n^2-4)A\RG}\)^{(n-2)/2}
-1\),
\eal
\eq57
$$
$$
\la_{c0}(\P_0)=6H_1^2=2\(\frac{m_{\u A}}{\mpl}\)^n
\frac{m_{\u A}^2}{\a_s^{2(n+1)}} \cdot
\frac{|\uP|^{(n+2)/2}n^{n/2}}{(n+2)^{(n+2)/2}\(\RG\)^{n/2}}.
\eq58
$$
The density $\rho_r$ is of course an \ef\ radiation density, because we
adsorbe into $\rho_r$ a factor with an \ef\ \gr\ \ct.

In our simplified model we get
$$
\al
\rho_r&=\La
\(\frac{98g^2(\z,\mu)+(\mu^2+4)\sqrt{49g^2(\z,\mu)+384h(\z,\mu)}
+32H(\z,\mu)}
{g^8(\z,\mu)(\mu^2+4)^2\cdot 2^{11}\cdot7^{14}}\right.\cr
&\qquad\qquad{}\times
\(7g(\z,\mu)+\sqrt{49g^2(\z,\mu)+384h(\z,\mu)}\)^6\cr
&\qquad\qquad{}\times\left.\vphantom{\frac{\sqrt{49g^2(\z,\mu)}}{(\mu^2+4)^2}}
\(\ln\(|\z|+\sqrt{\z^2+1}\)+2\z^2+1\)^6-1\) ,
\eal
\eq59
$$
$$
\al
\La&=\(\frac{m_{\u A}}{\mpl}\)^{14}\cdot\frac{m_{\u A}^2}{\a_s^{36}}
\cdot\frac{7^{14}}{2^{24}}\cr 
&\times\frac{g^8(\z,\mu)(\mu^2+4)^8}
{(\wiel)^7\(\ln\(|\z|+\sqrt{\z^2+1}\)+2\z^2+1\)^8}\,.
\eal
\eq60
$$
In the formulas \rf59--60 \ we should put $m_{\u A}=m\dr{EW}$ and
$\a_s^2=\a\dr{em}=\frac1{137}$. Let us notice that $\La$ is our contemporary
\co\ \ct\ given by \rf4 . It gives a scale for $\rho_r$.

The functions $g(\z,\mu)$, $h(\z,\mu)$, $H(\z,\mu)$ are given by Eqs (40),
(41a), (41b), (42) from Ref.~[1]. The numerical factor in front of~$\La$ can
be calculated  and we get
$$
\La=1.04\cdot 10^{-178}\,m\dr{EW}^2a(\z,\mu) \eq61
$$
where
$$
a(\z,\mu)=\frac{g^8(\z,\mu)(\mu^2+4)^8}{(\wiel)^7\(\lnz\)^8}\,. \eq62
$$
The calculated radiation density evolves in time according to the theory
developed in Ref.~[4].

In order to find some influence of $q_0$ (\q) field on the value of the \ef\
\gr\ \ct\ we consider a field equation for the scalar $q_0$ field in empty
space. One gets
$$
\(\frac{\pa^2q_0}{\pa t^2}-{\stpr\nabla}{}^2 q_0\)+
\t \e \o \a \exp(n\o\b q_0)\(\exp(2\o\b q_0)-1\)=0 \eq63
$$
where
$$
\ealn{
\o\a&=\frac{\la_{c0}n(n+2)\mpl^2}{16\pi\cdot\o M} &\rf64 \cr
\o\b&=\frac{\mpl}{2\sqrt{2\pi|\o M|}} &\rf65 \cr
\t\e&=\sgn\o M, \ \t\e{}^2=1.&\rf66
}
$$

Let us consider a static, spherically \sy ic case. In the spherical
coordinates one gets
$$
0=\frac1{r^2} \frac{d}{dr}\(r^2\frac{dq_0}{dr}\) - \t\e\o\a
\exp(n\o\b q_0)\(\exp(2\o\b q_0)-1\) \eq67
$$
where $q_0=q_0(r)$ is a function of $r$ only. In order to treat this equation
it is easier to come back to the old variable $\f=\o\b q_0$ (see Eq.~\rf28 ).
One gets
$$
\frac1{r^2} \frac{d}{dr}\(r^2\frac{d\f}{dr}\) - \frac14\t\e\o\la_{c0}
\exp(n\f)\(\exp(2\f)-1\)=0. \eq68
$$
We change the in\dc t variable $r$ into $\tau$
$$
\frac1{\tau^2} \frac{d}{d\tau}\(\tau^2\frac{d\f}{d\tau}\) - \t\e
\exp(n\f)\(\exp(2\f)-1\)=0 \eq69
$$
where
$$
r=\frac2{\sqrt{\o\la_{c0}}}\,\tau \eq70
$$
and
$$
\o\la_{c0}=\frac{n(n+2)\la_{c0}}{8\pi\o M}\,\mpl^2\,. \eq71
$$

We consider Eq.\ \rf69 \ in two regions:
\item{1)} for small fields $\f$,
\item{2)} for large fields $\f$.

In the first region we get
$$
\frac1{\tau^2} \frac{d}{d\tau}\(\tau^2\frac{d\f}{d\tau}\) + \t\e
e^{n\f}=0. \eq72
$$
In the second region we get
$$
\frac1{\tau^2} \frac{d}{d\tau}\(\tau^2\frac{d\f}{d\tau}\) - \t\e
e^{(n+2)\f}=0. \eq73
$$
Let us notice that both equations have similar nature and can be reduced to
the equation
$$
\frac1{x^2} \frac{d}{dx}\(x^2\frac{dy}{dx}\) +\e \t\e e^y=0 \eq74
$$
where in the first region $\e=1$,
$$
\ealn{
y&=n\f,&\rf75 \cr
x&=\sqrt n\,\tau, &\rf76
}
$$
and in the second region $\e=-1$,
$$
\ealn{
y&=(n+2)\f,&\rf77 \cr
x&=\sqrt {n+2}\,\tau, &\rf78
}
$$
We can transform \rf74 \ into
$$
x\,\frac{d^2y}{dx^2}+2\,\frac{dy}{dx}+\e\t\e xe^y=0 \eq79
$$
which is the celebrated Emden-Fowler equation known in the theory of gaseous
spheres (see~[5]).
Let us notice that the first region (small fields) means large distances and
the second region (large fields) means small distances. 

In this way we should consider Eq.~\rf79 \ in the region of small and
large~$x$. In the case of $\e\t\e=1$ the equation \rf74 \ has an exact
solution 
$$
y=\ln\(\frac2{x^2}\). \eq80
$$
Let us apply this to both regions (remembering that $\t\e$ in both cases has
a different sign).

One gets in the first region
$$
q_0=-\frac{2\sqrt{2\pi|\o M|}}{\mpl n} \ln\(\frac{r}{\frac{2\sqrt2}
{\sqrt{n\o\la_{c0}}}}\) \eq81
$$
and
$$
G\dr{eff}=G_N\(\frac{r}{\frac{2\sqrt2}
{\sqrt{n\o\la_{c0}}}}\)^{(n+2)/n}. \eq82
$$
In the second region
$$
q_0=-\frac{2\sqrt{2\pi|\o M|}}{\mpl (n+2)} \ln\(\frac{r}{\frac{2\sqrt2}
{\sqrt{(n+2)\o\la_{c0}}}}\) \eq83
$$
and
$$
G\dr{eff}=G_N\(\frac{r}{\frac{2\sqrt2}
{\sqrt{(n+2)\o\la_{c0}}}}\). \eq84
$$
In this way we get an interesting prediction for the behaviour of the
strength of \gr\ interactions.
In this very special solution $G\dr{eff}$ is going to zero if $r\to0$ and to
infinity if $r\to\infty$. 

Let us come to Eq.~\rf79 \ supposing $\e\t\e=1$. Thus we get
$$
x\,\frac{d^2y}{dx^2}+2\,\frac{dy}{dx}+ xe^y=0. \eq85
$$
Using an exact solution \rf80 \ we write
$$
y=y_1+\t y \eq86
$$
and consider Eq.~\rf85 \ for large $x$.

In this way we get an approximate solution (given by Chandrasekhar~[5])
$$
y=\ln\(\frac2{\eta^2}\)+\frac A{\sqrt\eta}\cos\(\frac{\sqrt7}2 \ln\eta\)
-2\ln\o\d, \quad |A|\ll 1, \eq87
$$
where
$\eta=\frac x{\,\o\d\,}$, $A$ and $\o\d$ are integration \ct s, $\o\d>0$. In this
way we get in the first region
$$
G\dr{eff}=\o G\dr{eff}\cdot\exp\(-\frac{\o A}{\sqrt \eta}\cos\(\frac{\sqrt7}
2 \ln\eta\)\) \eq88a
$$
where $\o A$ is a \ct\ ($|\o A|\ll 1$) and
$$
\eta=\frac{\sqrt{n\o\la_{c0}}\,\o\d}{2\sqrt2}\,r, \eq88b
$$
$\o G\dr{eff}$ is given by the formula \rf82 .

In the second region
$$
G\dr{eff}=\o{\o G}\dr{eff}\cdot\exp\(-\frac{\o{\o A}}{\sqrt \eta}\cos\(\frac{\sqrt7}
2 \ln\eta\)\)\eq88c
$$
where $\o{\o A}$ is a \ct\ ($|\o{\o A}|\ll1$), $\o{\o G}\dr{eff}$ is given by
the formula \rf84 \ and
$$
\eta=\frac{\sqrt{(n+2)\o\la_{c0}}\,\o\d}{2\sqrt2}\,r. \eq88d
$$
In this way we have very interesting non-Newtonian behaviour of $G\dr{eff}$
for large distances. Let us notice that the length scale is completely
arbitrary, because it is given by an integration \ct~$\o\d$. 

Let us consider Eq.\ \rf63 \ in Cartesian coordinates supposing flat \sy y
for a \q\ field $q_0=q_0(z,t)$ (nonstatic). One gets
$$
\(\frac{\pa^2q_0}{\pa t^2}-\frac{\pa^2q_0}{\pa z^2}\)
-\t \e \o \a \exp(n\o\b q_0)\(\exp(2\o\b q_0)-1\)=0. \eq89
$$
Let us change \dc t and in\dc t variables to $\xi,\eta,\f$:
$$
\ealn{
z&=\frac2{\sqrt{\o\la_{c0}}}\,\xi &\rf90 \cr
t&=\frac2{\sqrt{\o\la_{c0}}}\,\eta &\rf91 \cr
\f&=\o\b q_0. &\rf92 
}
$$
One gets
$$
\(\frac{\pa^2\f}{\pa \eta^2}-\frac{\pa^2\f}{\pa \xi^2}\)
-\t\e e^{n\f}\(e^{2\f}-1\)=0. \eq93
$$
Eq.\ \rf93 \ is an equation for flat scalar (\q) waves in our theory.
Let us consider it for large and small field~$\f$ (as before).

In this way one gets the equation
$$
\(\frac{\pa^2 y}{\pa T^2}-\frac{\pa^2y}{\pa x^2}\)
+\e\t\e e^y=0, \eq94
$$
where in the region of small field $\f$ we have $\e=1$ and
$$
\ealn{
y&=n\f &\rf95 \cr
x&=\sqrt n\, \xi &\rf96 \cr
T&=\sqrt n\, \eta &\rf97
}
$$
and in the region of large field $\f$, $\e=-1$ and
$$
\ealn{
y&=(n+2)\f &\rf98 \cr
x&=\sqrt {n+2}\, \xi &\rf99 \cr
T&=\sqrt {n+2}\, \eta. &\rf100
}
$$

Eq.\ \rf94 \ is the famous Liouville equation which can be transformed via a
B\"acklund \tr ation into a two-dimensional wave equation and afterwards
solved exactly. The general solution depends on two arbitrary functions $f$
and~$g$ of one variable, sufficiently regular. It is possible to consider
several problems for this equation: Cauchy initial problem, Darboux problem
and Goursat problem.

The general solution of \rf94 \ looks like ($\e\t\e=-1$)
$$
y(T,x)=\ln\[\frac{2g'(x-T)f'(x+T)}{\bigl(g(x-T)+f(x+T)\bigr)^2}\]
\eq101
$$
where $g'$ and $f'$ are derivatives of $g$ and $f$.

Thus one gets in the first region (small field)
$$
q_0(t,z)=\frac{2\sqrt{2\pi \o M}}{\mpl n}\cdot
\ln\[\frac{2g'\bigl(\frac{z-t}a\bigr)f'\bigl(\frac{z+t}a\bigr)}
{\(g\bigl(\frac{z-t}a\bigr)+f\bigl(\frac{z+t}a\bigr)\)^2}\], \eq102
$$
$$
a=\frac{2\sqrt2}{\sqrt{n\o\la_{c0}}}\,, \eq103
$$
and
$$
G\dr{eff}=G_N\(\frac{\(g\bigl(\frac{z-t}a\bigr)+f\bigl(\frac{z+t}a\bigr)\)^2}
{2g'\bigl(\frac{z-t}a\bigr)f'\bigl(\frac{z+t}a\bigr)}\)^{(n+2)/n}. \eq104
$$
In the second region (large field) 
$$
q_0(t,z)=\frac{2\sqrt{2\pi \o M}}{\mpl (n+2)}\cdot
\ln\[\frac{2g'\bigl(\frac{z-t}b\bigr)f'\bigl(\frac{z+t}b\bigr)}
{\(g\bigl(\frac{z-t}b\bigr)+f\bigl(\frac{z+t}b\bigr)\)^2}\], \eq105
$$
$$
b=\frac{2\sqrt2}{\sqrt{(n+2)\o\la_{c0}}}\,, \eq106
$$
and
$$
G\dr{eff}=G_N\cdot\frac{\(g\bigl(\frac{z-t}b\bigr)+f\bigl(\frac{z+t}b\bigr)\)^2}
{2g'\bigl(\frac{z-t}b\bigr)f'\bigl(\frac{z+t}b\bigr)}. \eq107
$$

In this way we get a spatio-temporal pattern of changing the \ef\ \gr\ \ct\
for small and large field regions. In both cases we have
$\e\t\e=-1$. However, in the small field region we have $\e=1$ and because of
this $\t\e=-1$. In the case of large field region $\e=-1$ and $\t\e=1$. In
order to be in line with our assumptions we should consider in the first case
such functions $f$ and~$g$ that the expression in~\rf105 \ is small and for
the second case vice versa.

Let us consider Eq.\ \rf63 \ in cylindrical coordinates supposing cylindrical
\sy y for the field~$q_0$, $q_0=q_0(\rho)$. One gets
$$
\frac1\rho\(\rho\,\frac{dq_0}{d\rho}\)-\e \o\a\exp(n\o\b q_0)\(\exp(2\o\b
q_0)-1\)=0. \eq108
$$
This equation can be \tr ed into
$$
\frac1\tau\,\frac{d}{d\tau}\(\tau\,\frac{d\f}{d\tau}\)-\t\e \exp(n\f)\(e^{2\f}
-1\)=0, \eq109
$$
$$
\rho=\frac2{\sqrt{\o\la_{c0}}}\,\tau. \eq110
$$
As usual we consider Eq.\ \rf109 \ in two regions for small and large fields.

In the first region
$$
\frac1\tau\,\frac{d}{d\tau}\(\tau\,\frac{d\f}{d\tau}\)+\t\e e^{n\f}=0. \eq111
$$
In the second region we get
$$
\frac1\tau\,\frac{d}{d\tau}\(\tau\,\frac{d\f}{d\tau}\)-\t\e e^{(n+2)\f}=0.
\eq112
$$
Both equations can be reduced to the equation
$$
\frac1x\,\frac{d}{dx}\(x\,\frac{dy}{dx}\)+\e\t\e e^y=0, \eq113
$$
where in the first region $\e=1$ and
$$
\ealn{
y&=n\f &\rf114 \cr
x&=\sqrt n\,\tau &\rf115 
}
$$
and in the second region $\e=-1$ and
$$
\ealn{
y&=(n+2)\f &\rf116 \cr
x&=\sqrt{n+2}\,\tau. &\rf117
}
$$
We can \tr\ \rf113 \ into
$$
x\,\frac{d^2y}{dx^2}+\frac{dy}{dx}+\e\t\e xe^y=0 \eq118
$$
which is the equation considered in~[6] for $\e\t\e=1$.

Following H. Lemke we write down a solution to Eq.\ \rf118 \ in a compact
form in three cases (concerning an integration \ct\ introduced by H.~Lemke).
We adopt his solutions to our problem.

\item{1)}$C=\k^2>0$, $\k$---arbitrary positive number:
$$
y(x)=-2\ln\(\frac x{\,\o a\,}\(\(\frac{\o a}{x}\)^\k +\(\frac{x}{\,\o a\,}\)^\k \)\)
+\ln\(4\k^2\o a{}^2\) \eq119
$$
where $\o a>0$ is an arbitrary \ct.

\item{2)}$C=-\om^2<0$, $\om$ is an arbitrary positive number:
$$
y(x)=-2\ln\(2x\sin(\om \ln x + \d)\)+\ln(4\om^2) \eq120
$$
where $\d$ is an arbitrary \ct.

\item{3)}$C=0$:
$$
y(x)=-2\ln\(\frac x{\,\o a\,}\ln\(\frac x{\,\o a\,}\)\)-2\ln\o a \eq121
$$
where $\o a>0$ is an arbitrary \ct.

Using these solutions we write down a spatial \dc ce of $G\dr{eff}$ in the
case of small and large fields. 
In the case of small fields one gets
$$
\dsl{
\tin1 \hfill G\dr{eff}=G_N\(4\k^2\o a{}^2\)^{(n+2)/n}
\frac{\(1+r^{2n}\)^{2(n+2)/n}}{r^{2(\k-1)(n+2)/n}} \hfill \rf122
}
$$
where
$$
r=\frac1{2\o a}\sqrt{n\o\la_{c0}}\,\rho. \eq123
$$
$$
\dsl{
\tin2 \hfill G\dr{eff}=\frac{G_N(\o\d)^{2(n+2)/n}}{(\om^2)^{(n+2)/n}}
\, r^{2(n+2)/n}\bigl(\sin(\om\ln r)\bigr)^{2(n+2)/n} \hfill \rf124 
}
$$
where
$$
\ealn{
&r=\frac1{2\o\d}\sqrt{n\o\la_{c0}}\,\rho &\rf125 \cr
&\om\ln\o\d=-\d &\rf126
}
$$
$$
\dsl{
\tin 3\hfill G\dr{eff}=G_N\o a^{2(n+2)/n}r^{2(n+2)/n}(\ln r)^{2(n+2)/n} \hfill
\rf127 
}
$$
and $r$ is given by Eq.\ \rf123 .

In the case of large field one gets:
$$
\dsl{
\tin1 \hfill G\dr{eff}=G_N(4\k^2\o a^2)\frac{(1+r^{2\k})^2}{r^{2(\k-1)}} \hfill
\rf128 
}
$$
where
$$
r=\frac1{2\o a}\sqrt{(n+2)\o\la_{c0}}\,\rho. \eq129
$$
$$
\dsl{
\tin2 \hfill G\dr{eff}=G_N\(\frac{\o\d}{\om}\)^2 r^2 \bigl(\sin(\om\ln r)\bigr)^2
\hfill \rf130
}
$$
where
$$
r=\frac1{2\o\d}\sqrt{(n+2)\o\la_{c0}}\,\rho , \quad \ln \o\d=-\frac\d\om. \eq131
$$
$$
\dsl{
\tin3 \hfill G\dr{eff}=G_N\o a^2r^2(\ln r)^2 \hfill \rf132
}
$$
and $r$ is given by Eq.\ \rf129 . 

Let us notice that in that spatial \dc ce
for large and small field we have $\k$ and $\o a$ ($\o\d$, $\om$) as
integration \ct s. In this way integration \ct s induce a power law of this
\dc ce and also a scale. For sufficiently big~$n$ ($n>14$) there is no
significant difference between both cases. It means the \q\ field behaves
everywhere as for large field case (in these solutions of course). It is
evident that the spatial \dc ce (in cylindrical \sy y case) of~$G\dr{eff}$
goes to some kind of the fifth force. However, we have to do not with a
universal law of Nature but rather with some kind of initial conditions. Thus
the real \dc ce of $G\dr{eff}$ on spatial coordinates can be obtained after
averaging~$G\dr{eff}$ \wrt initial conditions. Let us describe it using
Eq.~\rf128 .

Let us write $G\dr{eff}$ in a form where $\k$ and $\o a$ are explicitly
visible:
$$
\ealn{
G\dr{eff}&=G_N\(4\k^2\o a^2\)\cdot
\frac{\(1+\(\frac x{\,\o a\,}\)^{2\k}\)^2}{\(\frac x{\,\o a\,}\)^{2(\k-1)}},
&\rf133 \cr
x&=\frac12 \sqrt{(n+2)\o\la_{c0}}\,\rho. &\rf134 
}
$$
The \ct s $\o a$ and $\k$ are chosen randomly for they are \dc t on initial
conditions. Let $\mu$ be a measure defined on $(0,+\infty)^2$, positive and
normalized to~1, i.e.
$$
\intop_0^{+\infty} \,\intop_0^{+\infty} d\mu(\o a,\k)=1. \eq135
$$
This measure gives an account how frequently we have to do with some initial
conditions (i.e.\ with integration \ct s $\o a$,~$\k$). In this way an
experimental \dc ce of~$G\dr{eff}$ on~$x$ should be obtained from
$$
E(G\dr{eff})=4G_N \intop_0^{+\infty}\, \intop_0^{+\infty} 
\frac{\k^2\o a^2\(1+\(\frac x{\,\o a\,}\)^{2\k}\)^2}
{\(\frac x{\,\o a\,}\)^{2(\k-1)}}\,d\mu. \eq136
$$
It means it is an expectation value of $G\dr{eff}$ \wrt the measure~$\mu$.
$\mu$~need not be absolutely continuous \wrt the Lebesgue measure on ${\Bbb
R}^2$. The formula \rf136 \ could give an account on some serious problems
with comparisons of several measurements of~$G$ (a~Newton \ct). Maybe we have
to do with different initial conditions for \q\ field. These initial
conditions appear with different probabilities according to the measure
$d\mu(\o a,\k)$ going to an expectation value $E(G\dr{eff})$. The second
central moment of $d\mu(\o a,\k)$ (if it exists) can express a deviation from
the law described by $E(G\dr{eff})$. 

In the simplest case we can suppose a continuous Gaussian distribution
for~$\k$ only:
$$
d\mu(\k)=\frac1{\sqrt{2\pi}\,\s}\,e^{-(\k-\k_0)^2/(2\s^2)}\,d\k.
$$
In this case we have
$$
\o G\dr{eff}(x)=E(G\dr{eff}(x))=
\frac{G_N\o a^2}{\sqrt{2\pi}\,\s} \intop_{-\infty}^{+\infty}
d\k\,e^{-(\k-\k_0)^2/(2\s^2)}\k^2\,\frac{\(1+\(\frac x{\,\o a\,}\)^{2\k}\)^2}
{\(\frac x{\,\o a\,}\)^{2(\k-1)}}\,. \eq136a
$$
One gets after some algebra
$$
\al
\o G\dr{eff}(x)&=G_N \k_0 \exp\(\o a-\frac{\k_0^2}{2\s^2}\)\(\frac x{\,\o a\,}\)
^2\cr
&\times\Biggl[4\s^4\exp\frac{\(2\ln\(\frac x{\,\o a\,}\)\s^2+\k_0^2\)^2}{2\s^2}
\ln\(\frac x{\,\o a\,}\)\cr
&\qquad{} +(2\s^2+\k_0^4)
\exp\frac{\(2\ln\(\frac x{\,\o a\,}\)\s^2+\k_0^2\)^2}{2\s^2}\cr
&\qquad{}-4\s^4\exp\frac{(\s+\k_0)^2}2 \ln^2\(\frac x{\,\o a\,}\)+
2(\s^2+\k_0^2)\exp\frac{(\s+\k_0)^2}2 \Biggr].
\eal
\eq136b
$$

In the case of the solution \rf107 \ parametrized by two functions
of one variable we can consider them as random variables parametrized by~$z$
and~$t$. Moreover in this case we should consider a measure~$\mu$ on an
infinite-dimensional space of functions $f$ and~$g$, supposing that solution
\rf107 \ is generalized to this space. Afterwards we can use as~$\mu$ the
Wiener or Gaussian measure in $L^2$ space. If we use for Gaussian
distribution the normalized distribution $N(0,1)$, the formula \rf136b \
simplifies to
$$
\al
\o G\dr{eff}(x)&=e^{\bar a}\sqrt e \, G_N\cr
&\times\[4\(\frac x{\,\o a\,}\)^{2(1+\ln(x/\bar a))}\ln\(\frac x{\,\o a\,}\)
+\(\frac x{\,\o a\,}\)^{2(1+\ln(x/\bar a))} -4\ln^2\(\frac x{\,\o a\,}\)+2\]. 
\eal
\eq136c
$$

Let us consider three our cases \rf119 , \rf120 \ and \rf121 \ for small and
large fields cases.

The small field case is such that
$$
e^\f<1, \qquad\text{i.e. }\f<0. \eq137
$$
The large field case is if
$$
e^\f>1, \qquad\text{i.e. }\f>0. \eq138
$$
For \rf119 \ we have for the small case
$$
f(r)>1,
$$
where
$$
f(r)=r^{1-\k}+r^{\k+1}, \qquad r=\(\frac x{\,\o a\,}\). \eq139
$$

Let us consider two cases
$$
0<\k<1 \qquad\text{and}\qquad \k>1.
$$
In the first case $f(r)\nearrow$ in $[0,+\infty)$ and we have simply
$r>r_0$ where $r_0$ satisfies the equation
$$
r_0^{2\k+1}+r_0-1=0. \eq140
$$
In the second case $\k>1$,
$$
\lim_{r\to0} f(r)=+\infty \qquad\text{and}\qquad \lim_{r\to\infty}
f(r)=+\infty  .
$$
The function $f(r)$ has a minimum at
$$
r_1=\(\frac{\k-1}{\k+1}\)^{1/\k}. \eq141
$$
Let us calculate $f(r_1)$.
$$
f(r_1)=\(\frac{\k+1}{\k-1}\)^{\k-1/\k}+\(\frac{\k-1}{\k+1}\)^{\k+1/\k}. \eq142
$$
It is easy to see that if $\k>1$, then $\k-1/\k>0$. This means that
$$
f(r_1)>1 \eq143
$$
and therefore $f(r)>1$ for all $r>0$. Thus simultaneously we get a solution
for large field only if $\k<1$, i.e.
$$
r<r_0. \eq144
$$
If $\k=1$, we always have $f(r)\ge1$ (i.e.\ only a small field).

Let us consider \rf120 . In this case the small field condition reads
$$
h(r)=r\sin(\om\ln r)>1, \quad \text{where }r=\frac x {\,\o \d\,}\,. \eq145
$$
First of all we need $h(r)>0$. Let us observe that $|h(r)|\le r$ for
every~$r>0$. Next we see that the roots of $h(r)$ are the numbers
$$
r_{0,k}=e^{k\pi/\om}, \quad k=0,\pm1,\pm2,\ldots \eq146a
$$
and
$$
h(r)>0 \quad\text{if}\quad 
r_{0,2k}<r<r_{0,2k+1},\quad k=0,\pm1,\pm2,\ldots \eq146b
$$
The maximum of the function $h(r)$ in the interval
$(r_{0,2k},r_{0,2k+1})$ is greater than
$$
h(r_{0,2k+1/2})=e^{2k\pi/\om}\cdot e^{\pi/(2\om)}, \eq147
$$
but smaller than $r_{0,2k+1}$. Thus the maxima are smaller than~1 for
negative $k$ and greater than~1 if $k=0,1,2,\ldots$. It
means that in each interval $(r_{0,2k},r_{0,2k+1})$, $k=0,1,2,\ldots$, there
exist two numbers $r_{3,2k}$ and $r_{2,2k}$ such that $r_{3,2k}<r_{2,2k}$,
$$
h(r_{3,2k})=h(r_{2,2k})=1 \eq148
$$
and
$$
h(r)>1 \quad\text{if}\quad r_{3,2k}<r<r_{2,2k}\,. \eq149
$$

The condition for large fields, $0<h(r)<1$, is satisfied if
$$
r_{0,2k}<r<r_{0,2k+1}, \quad k=-1,-2,\ldots \eq150
$$
or
$$
r_{0,2k}<r<r_{3,2k}, \quad \text{or}\quad r_{2,2k}<r<r_{0,2k+1},
\quad k=0,1,2,\ldots \eq151
$$

In the third case, i.e.\ Eq.\ \rf121 , one has for the small field case
$$
r \ln r>1 \eq152
$$
where $r=\frac x{\,\o a\,}$. Let $r_4\ln r_4=1$, $r_4>1$. We have
$r>r_4=1.7632\ldots$. In the large field case 
$$
0<r<r_4=1.7632\ldots. \eq153
$$

Thus we have in general large fields on large distances.

In this way we have solutions for large and small distances. One can try to
connect them to get a solution for all distances. However in this case it is
necessary to be very careful, for our solutions depend on some integration
\ct s which can be different for both asymptotic regions.

Let us consider Eq.\ \rf63 \ in two special cases:

\item{I} $q_0=q(z)$ --- static and depending only on~$z$;
\item{II} $q_0=q_0(t)$ --- non-static and spatially \ct.

Let us consider also these cases for small and large fields~$\f$. In all of
these cases we come to the following equation
$$
\frac{d^2y}{dx^2} + \e_1\e\t\e e^y=0 \eq154
$$
where $\e_1=1$ for case I and $\e_1=-1$ for case II. 

Eq.\ \rf154 \ can easily be reduced to the integral
$$
x-x_0=\frac{\e_3}{\sqrt2} \int \frac{dy}{\sqrt{\e_2\om^2-\eta e^y}} \eq155
$$
where $\eta=\e_1\e\t\e$, $\eta^2=1$, $C=2\e_2\om^2$, $\om\ge0$, $\e_2^2=1$ is
an integration \ct, $\e_3^2=1$ and $x_0$ also is an integration \ct.

After some calculation we get the following solutions:
$$
\dsl{
\text{A.}\hfill y(x)=2\[\ln \om-\ln\left|\sinh\(
\frac{\sqrt2\,(x-x_0)\om}2 \)\right|\], \quad \eta=-1,\ \e_2=1\hfill \rf156 \cr
\text{B.}\hfill y(x)=2\[\ln \om-\ln\left|\sinh\(
\frac{\sqrt2\,(x-x_0)\om}2 \)\right|\], \quad \eta=1,\ \e_2=1\hfill \rf157
}
$$
where
$$
\frac{\sqrt2\,(x-x_0)\om}2 >\ln(1+\sqrt2)\eq158 
$$
or
$$
\frac{\sqrt2\,(x-x_0)\om}2 <\ln(\sqrt2-1).\eq159
$$
$$
\dsl{
\text{C.}\hfill y(x)=2\[\ln\om-\ln\left|
\cos\(\frac{\sqrt2\,(x-x_0)\om}2\)\right|\], 
\quad \eta=-1,\ \e_2=-1. \hfill \rf160
}
$$

Let us apply these solutions to our problems. First of all let us consider a
static configuration with $z$ \dc ce only. In this case $\e_1=1$ and $\eta=\e
\t\e$. For the small field case one gets, $\e=1$, $\eta=\t\e$,
$$
G\dr{eff}=\frac{G_N}{\om^{2(n+2)/n}}\left|\sinh p\right|^{(n+2)/n} \eq161
$$
where
$$
p=\frac{\om(z-z_0)}4 \,\sqrt{n\o\la_{c0}}\,. \eq162
$$
For the large field case we get, $\e=-1$, $\eta=-\t\e$,
$$
G\dr{eff}=\frac{G_N}{\om^2}\left|\cos p\right| \eq163
$$
where
$$
p=\frac{\om(z-z_0)}4 \,\sqrt{(n+2)\o\la_{c0}}\,. \eq164
$$
In this case $\t\e=1$ for $\eta=-1$.

In a nonstatic configuration $\e_1=-1$ and $\eta=-\e\t\e$. For the small
field case ($\e=1$), $\eta=-\t\e$,
$$
G\dr{eff}=\frac{G_N}{\om^{2(n+2)/n}}\left|\sinh q\right|^{(n+2)/n} \eq165
$$
where
$$
q=\frac{\e_3\om(t-t_0)}4 \sqrt{n\o\la_{c0}} \eq166
$$
and analogically for the large field case ($\e=-1$), $\eta=\t\e$,
$$
G\dr{eff}=\frac{G_N}{\om^2}\left|\cos q\right|, \eq167
$$
$$
q=\frac{\om(t-t_0)}4 \,\sqrt{(n+2)\o\la_{c0}}\,. \eq168
$$
In this case $\t\e=-1$ for $\eta=-1$.

Let us notice that in a static configuration for small field we have two
possibilities for $\eta=-1$ (no condition on~$p$) and $\eta=1$ (conditions
\rf158--159 ). Thus without conditions we have $\t\e=-1$ and with conditions
$\t\e=1$. In a non-static configuration for small field we have vice versa
$\t\e=1$ without conditions and $\t\e=-1$ with conditions \rf158--159 .

Let us come back to the Eq.\ \rf89 \ and consider it in a travelling wave
scheme. In this way we have
$$
q_0(z,t)=\t q(z-vt) \eq169
$$
where $v$ is a velocity of the travelling wave (a soliton), $|v|<1$. Let us
consider this equation in both regimes (for small and large fields). In this
way we come to the expression
$$
(1-v^2)\,\frac{d^2\chi}{d\xi^2}-\e\t\e e^\chi=0 \eq170
$$
where $\chi$ is a shape function of a soliton. Changing an independent
variable from $\xi$ to~$\la$ one gets
$$
\frac{d^2\chi}{d\la^2}-\e_1\e\t\e e^\chi=0 \eq171
$$
where $\e_1=-1$, i.e.\ we get Eq.\ \rf154 \ with $\eta=-\e\t\e$,
$$
\la=\frac \xi {\sqrt{1-v^2}},\quad \xi=\sqrt{1-v^2}\,\la. \eq172
$$

In this way we adopt our solutions A, B, C in both regimes: small and large
field (changing $\chi$ into~$\la$). For small field we get ($\e=1$,
$\eta=-\t\e$) 
$$
q_0(z,t)=\frac2{\o \b n}\[\ln\om - \ln\left|\sinh p\right|\] \eq173
$$
where
$$
p=\frac{\om\sqrt{n\o\la_{c0}}}{4\sqrt{1-v^2}}\,(z-vt). \eq174
$$
For large field we get ($\e=-1$, $\eta=\t\e$) 
$$
q_0(z,t)=\frac2{\o \b (n+2)}\[\ln\om - \ln\left|\cos p\right|\] \eq175
$$
where
$$
p=\frac{\om\sqrt{(n+2)\o\la_{c0}}}{4\sqrt{1-v^2}}\,(z-vt). \eq176
$$
For \rf173 \ we have $\eta=-\t\e$ and because of this $\t\e=1$ without any
conditions and if $\t\e=-1$ we have conditions \rf158--159 . In the case of
the formula \rf174 \ $\eta=-1$ and $\t\e=-1$.

We can write down formulas for $G\dr{eff}$ in the soliton case
$$
G\dr{eff}=\frac{G_N}{\om^{2(n+2)/n}}\left|\sinh p\right|^{(n+2)/n} \eq177
$$
and $p$ is given by the formula \rf174 \ (with or without conditions
\rf158--159 ). This is of course a small field case.

In the large field case
$$
G\dr{eff}=\frac{G_N}{\om^2}\left|\cos p\right|, \eq178
$$
and $p$ is given by the formula \rf176 . In this case $\t\e=-1$. (Let us
notice that this is a case of SO(3) group in our theory.)

Let us notice that conditions \rf158--159 \ can be considered as conditions
for small field in $z$ or $t$ domains. Let us notice that in our solutions
concerning a behaviour of an \ef\ \gr\ \ct\ we get completely arbitrary
length or time scale (given by integration \ct s). In this way a spatial or
time \dc ce of $G\dr{eff}$ can be (except the solution \rf80 \ and
simultaneously the approximate solution in the case of spherical \sy y) such
that $G\dr{eff}$ can be really \ct\ on distances (or times) accessible in
experiments. Only a statistical approach mentioned here can give a light on
this \dc ce to compare it with an experiment.

In order to connect our results to the ordinary \gr\ physics we consider
again Eq.~\rf67 \ in small field regime for initial conditions $\f(0)=0$ and
$\frac{d\f}{dx}(0)=0$. The first condition means that we want to have
$G\dr{eff}(0)=G_N$ and the second that the \q\ field does not grow quickly.
The problem cannot be solved analytically. Moreover R.~Emden in the first
point of Ref.~[5] did it for us. We quote here his results adopted to our
notation ($\e=+1$ and $\t\e=-1$).

\def\nh{\cr\noalign{\hrule}}
$$\vbox{\offinterlineskip
\halign{\strut\vrule\quad \hfil$#$\quad &\vrule\quad \hfil$#$\quad 
&\vrule\quad $#$\hfil \quad&\vrule\quad \hfil$#$\quad \hfil\vrule\cr
\noalign{\hrule}
\quad x\hfil&\quad -y\hfil&\hfil e^{y}\quad &\quad G\dr{eff}/G_N\nh
0.00&0.00000&1.00000&1.00000\nh
0.25&0.01037&0.98969&0.98823\nh
0.50&0.04113&0.95971&0.95409\nh
0.75&0.09113&0.91290&0.90109\nh
1.00&0.15903&0.85296&0.83380\nh
1.25&0.24225&0.78486&0.75816\nh
1.50&0.33847&0.71285&0.67920\nh
1.75&0.44488&0.64090&0.60143\nh
}}
$$
$$\vbox{\offinterlineskip
\halign{\strut\vrule\quad \hfil$#$\quad &\vrule\quad \hfil$#$\quad 
&\vrule\quad $#$\hfil \quad&\vrule\quad \hfil$#$\quad \hfil\vrule\cr
\noalign{\hrule}
\quad x\hfil&\quad -y\hfil&\hfil e^{y}\quad &\quad G\dr{eff}/G_N\nh
2.00&0.55967&0.57140&0.52749\nh
2.50&0.80584&0.44671&0.39813\nh
3.00&1.06226&0.34537&0.29670\nh
3.50&1.31937&0.26730&0.22138\nh
4.00&1.57071&0.20790&0.16611\nh
4.50&1.81246&0.16325&0.12601\nh
5.00&2.04264&0.12968&0.09686\nh
6.00&2.46598&0.08493&0.05971\nh
7.00&2.84160&0.05833&0.03887\nh
8.00&3.17489&0.04180&0.02656\nh
9.00&3.47128&0.03108&0.01893\nh
10.00&3.73646&0.02384&0.01398\nh
100&8.59506&0.000175&5.0854\cdot10^{-5}\nh
1000&13.09847&0.000002&3.0683\cdot10^{-7}\nh
}}
$$
where
$$
\ealn{
&x=\frac12 \sqrt{n\o\la_{c0}}\,r \cong \(\frac r{10\,\text{Mpc}}\),&\rf179 \cr
&\frac{G\dr{eff}}{G_N}=\(e^{y}\)^{(n+2)/n}=\(e^{y}\)^{8/7}. &\rf180
}
$$
We take $n=14$.

It is easy to see that for large $n$
$$
\frac{G\dr{eff}}{G_N}=e^{y}. \eq181
$$

It is easy to see that on a distance of $1\,$Mpc $G\dr{eff}$ does not differ
from~$G_N$. Even on a distance of $10\,$Mpc it is about 10\% smaller. Thus in
the Solar System Newtonian \gr\ physics does not change. Even on the level of
a galaxy this change is minimal and cannot be observed. Moreover, there is an
important conclusion: on distances about $200\,$Mpc the strength of \gr\
interactions is about $10^{-5}$ times this on short distances measured in the
Solar System (and for $10^3$\,Mpc of $10^{-7}$). It is hard to tell how it influences a mass of a cluster of
galaxies if we realize that from any observational data only a product $GM$
has been obtained (not~$M$).

From the other side on distances of $100\,$Mpc the strength of \gr\
interactions is very weak (not only because of the distance). Thus if we
consider clusters of galaxies as substrat \pc s in cosmology then they do not
interact.

Let us consider Eq.\ \rf63  \ in Cartesian coordinates for two-dimensional
static case (i.e. $\frac{\pa}{\pa z}=0$, $\frac{\pa}{\pa t}=0$). One gets
$$
\(\frac{\pa^2q_0}{\pa x^2}+\frac{\pa^2q_0}{\pa y^2}\)
-\t \e \o \a \exp(n\o\b q_0)\(\exp(2\o\b q_0)-1\)=0 \eq182
$$
(where $\o\a$, $\o\b$ are given by formulas \rf64 \ and \rf65 ).

As usual, we come to the formula
$$
\(\frac{\pa^2\f}{\pa x_1^2}+\frac{\pa^2\f}{\pa x_2^2}\) - \t\e e^{n\f}
(e^{2\f}-1)=0 \eq183
$$
where
$$
\left.\matrix x\\y\endmatrix\right\} = \frac2{\sqrt{\o \la_{c0}}}\,x_i,
\qquad i=1,2. \eq183a
$$
We consider Eq.\ \rf183 \ for small and large fields and we get
$$
\(\frac{\pa^2\chi}{\pa z_1^2}+\frac{\pa^2\chi}{\pa z_2^2}\) - \e\t\e e^{\chi}
=0 \eq184
$$
where as usual for a small field $\e=1$ and
$$
\ealn{
\chi&=n\f, &\rf185 \cr
z_i&=\sqrt n\,x_i &\rf186
}
$$
and for a large field $\e=-1$ and
$$
\ealn{
\chi&=(n+2)\f, &\rf187 \cr
z_i&=\sqrt {n+2}\,x_i\,. &\rf188
}
$$
Thus we come to the equation known as Liouville equation
$$
\D \chi=e^\chi \eq184a
$$
if $\e\t\e=1$.

This equation can be explicitly solved. First of all we change in\dc t
variables into
$$
\ealn{
Z&=\frac1{\sqrt2}(z_1+iz_2) &\rf189 \cr
\chi(Z)&=-\ln\(\tfrac12(1-|g|^2)\)+\tfrac12 \ln\left|\frac{dg}{dZ}\right|
&\rf190
}
$$
where $g$ is an arbitrary analytic function on a complex plane $Z$.

In this way we get for the small field case
$$
G\dr{eff}=G_N\(\frac{1-|g(Z)|^2}2\)^{(n+2)/n}
\(\left|\frac{dg}{dZ}(Z)\right|\)^{-(n+2)/(2n)} \eq191
$$
where
$$
Z=\sqrt{\frac{\o\la_{c0}}{2n}}\cdot(x+iy). \eq192
$$

In the large field case
$$
G\dr{eff}=G_N\(\frac{1-|g(Z)|^2}2\)
\(\left|\frac{dg}{dZ}(Z)\right|\)^{-1/2} \eq193
$$
where
$$
Z=\sqrt{\frac{\o\la_{c0}}{2(n+2)}}\cdot(x+iy). \eq194
$$
Eqs \rf191 \ and \rf193 \ can have very interesting behaviour for
$\frac{dg}{dZ}$ could have some singularities. The physical interpretation of
these singularities can be very interesting.

Similarly as for Eqs \rf104 \ and \rf107 \ we can consider an expectation
value of $G\dr{eff}$ \wrt some kind of normalized measure for a space of
analytic functions on the complex plane (respectively chosen).

It seems that an assumption all the energy of a \q\ is stored as \q\ \pc s is
unrealistic. Let us suppose that only a fraction of this energy is stored as
\pc s. Let this fraction be $\eta$,
$$
0<\eta<1. \eq195
$$
$\eta$ can be a function of time and even of a space-point (locally).

Let us suppose that a gas of \q\ \pc s is a perfect gas governed by the
Clapeyron equation. Thus we have
$$
\frac{p_1}{\rho_1}=\frac{K_BT}{m_0}=\frac T{T_0}, \quad
T_0=\frac{m_0}{K_B}\simeq 0.11\,{}^\circ\text{K} \eq196
$$
(if we take $m_0\simeq 10^{-5}\,$eV). $T$ is the temperature of a gas.

One gets
$$
\ealn{
\rho_1&=\eta\rho_Q=\eta\rho &\rf197 \cr
p&=p_Q+p_1=p_Q+\eta\,\frac T{T_0}\, \rho=
-\rho(1-\eta)+\rho\,\frac T{T_0}\,\eta. &\rf198
}
$$
Eventually one gets
$$
\ealn{
p&=\rho\(\eta\(1+\frac T{T_0}\)-1\) &\rf199 \cr
p&=w\rho &\rf200 \cr
w&=\eta\(1+\frac T{T_0}\)-1. &\rf201 
}
$$

Now we can calculate an isothermic speed of sound in a \q.
$$
C_1^2=\frac p\rho=w=\eta\(1+\frac T{T_0}\)-1. \eq202
$$
For
$$
0<C_1^2<1 \eq203
$$
we get
$$
\frac{T_0}{T+T_0}<\eta<\min\[\frac{2T_0}{T_0+T},1\]. \eq204
$$
Let us remind to the reader that an isothermic sound is appropriate for low
frequency of acoustic waves (we have to do with this sound in astrophysics).
In general we have to do with so called adiabatic sound. In order to
calculate a speed of an adiabatic sound in a \q\ we should find an analogue
for a Poisson adiabate for our equation of state. One gets supposing that an
internal energy of a \q\ is an energy of one-atomic gas of \q\ \pc s
$$
dU+p\,dV=0 \eq205
$$
or
$$
\frac32\,dT=\frac{d\rho}{\rho}\(\eta-1+\frac T{T_0}\,\eta\) \eq206
$$
and finally
$$
\ealn{
&\frac p{\rho^\k}=\text{const.} &\rf207 \cr
&\k=\frac{2\eta+3}3 &\rf208
}
$$
and a speed of an adiabatic sound simply reads
$$
C_2=\sqrt{\frac{\k p}{\rho}}= \sqrt{\k\(\eta\(1+\frac T{T_0}\)-1\)}\,. \eq209
$$
It is interesting to ask what kind of a polythrope $\k$ represents. Let us
remind to the reader that in general
$$
\k=\frac{c_p-c}{c_v-c} \eq210
$$
where $c$ is a specific heat of a polythrope in mind.

One gets in our case
$$
c=\frac{3(\eta-1)}{2\eta}\,\o R \eq211
$$
where $\o R$ is a universal gas \ct.

Thus we found both speeds of sound in a \q\ for low frequency and for
high frequency of sound.
The measurement of both speeds can help us to find $\eta$ and $T$. One
gets 
$$
\eta=\frac32\(\(\frac{C_2}{C_1}\)^2-1\) \eq212
$$
and
$$
T=\frac{T_0}{3(C_2^2-C_1^2)}\(C_1^2(2C_1^2+5)-3C_2^2\). \eq213
$$
Let us notice that a gas of \q\ \pc s is quite cold.
Moreover these \pc s are highly relativistic. For the temperature
$T\simeq T_0=0.11\,{}^\circ$K one sees that a speed of a \q\ \pc \ is about
0.91 of the speed of light. Moreover we can consider lower temperatures.
It is interesting to notice that the speed of sound is of the same order.

Thus if we want to have both speeds smaller than a speed of light,
$$
0<C_i^2<1, \quad i=1,2, \eq214
$$
we get
$$
\eta>\frac{1+\sqrt{13+12t}}{1+t}=f(t) \eq215
$$
where
$$
t=\frac T{T_0}\,. \eq216
$$
The condition
$$
t=\frac T{T_0}>13 \eq217
$$
guarantees that $f(t)<1$. Moreover for $t=14$ one gets
$$
\eta\ge 0.96357. \eq218
$$
This seems quite interesting, however maybe too much. Let us calculate a mean
scattering length of \q\ \pc s for such a big $\eta=0.96357$
$$
l\dr{scattering}=\frac1{\eta\s n}=1.0378\cdot l. \eq219
$$
We have still to do with a very dense gas of \q\ \pc s.

It seems that it would be reasonable to repeat these calculations using a
different equation of state for \q\ \pc s gas. In particular we can consider
a gas of \q\ \pc s as a massless boson gas (with spin zero). In this case the
equation of state looks:
$$
p=p_Q+p_1=-(1-\eta)\rho+\frac{4\eta\s\rho}3\,T^4 \eq220
$$
and an adiabate equation
$$
dU=\o c_v\,dT+p\,dV=0 \eq221
$$
where
$$
\o c_v=16\s T^3 , \eq222
$$
$\s$ is the Stefan-Boltzmann \ct.

One easily integrates
$$
\frac p{\rho^{\bar\k}}=\text{const.} \eq223
$$
where
$$
\o\k=\frac{3+\eta}3\,. \eq224
$$
We have as before two kinds of sound: an isothermic sound
$$
\o C_1=\sqrt{\frac p\rho}=\sqrt{\eta-1+\frac{4\eta\s}3\,T^4} \eq225
$$
and an adiabatic sound
$$
\o C_2=\sqrt{\o\k\(\eta-1+\frac{4\eta\s}3\,T^4\)}=\sqrt{\frac{\o\k p}\rho}\,. 
\eq226
$$

From $0<\o C_i^2<1$, $i=1,2$, one gets
$$
\frac1{1+\(T/{\,\o T_0\,}\)^4} < \eta < \min\[1,\frac2{1+\(T/{\,\o
T_0\,}\)^4} \] \eq227
$$
and
$$
\eta>\frac{-3t^4-1+\sqrt{9t^8+18t^4+13}}{2(1+t^4)}=\o f(t) \eq228
$$
with the condition
$$
t=\frac T{\,\o T_0\,}>\frac1{\sqrt2}\simeq 0.707. \eq229
$$
This condition guarantees that
$$
\o f(t)<1 \eq230
$$
where $T_0^4=\frac3{4\s}$ or
$$
\o T_0=\frac1{K_B}\root4\of{\frac{360}\pi}\simeq\frac{3.27}{K_B}\,, \eq231
$$
$K_B$ is the Boltzmann \ct.

From the measurements of $\o C_1$ and $\o C_2$ we can obtain as before $\eta$ and
$T$. One gets:
$$
\eta=3\(\(\frac{\o C_2}{\,\o C_1\,}\)^2-1\) \eq232
$$
and
$$
T=\frac{\o T_0}{\root4\of3}\(\frac{4\o C_1^2+\o C_1^4-3\o C_2^2}{\o C_2^2-\o
C_1^2}\)^{1/4}. \eq233
$$
Recently many papers have appeared concerning the speed of sound in a \q\
(see Ref.~[7]). In some of them the authors propose to measure this speed.

There are interesting propositions to include primordial \gr\ waves to
fluctuations of a \q\ and vice versa. We examine these problems below.

We consider a mass of a \q\ \pc, a speed of sound in a \q\ and
several solutions to \q\ equations coming to the interesting behaviour of an
\ef\ \gr\ \ct. Some of issues of our theory are considered in
self-interacting Brans-Dike theory [8].

The fraction $\eta$ can be connected with the part of an energy density of
\q\ field $Q$ in such a way that it is a fraction of fluctuation energy density
around an equilibrium~$Q_0$. In this way the $q_0$ field which can evolve due
to an evolution of~$Q$ and due to fluctuations of primordial \gr\ waves will
be a source of a gas of thermalized \pc s. It seems that an approach with a
boson equation of state is more appropriate to consider. 

It is interesting to consider a fraction of a dark energy stored as boson
particles as a dark matter in the \U.

Let us consider a primordial spectrum of \gr\ waves in our approach. This
spectrum is flat for our Hubble \ct\ is really \ct
$$
P\dr{grav}=\frac2{M^2\dr{pl}}\(\frac{H_0}{2\pi}\)^2 \eq235
$$
(see Ref.\ [9]). Thus
$$
n\dr{gr}=1. \eq236
$$

Now we start to examine \q\ fluctuations caused by fluctuations of a metric.
In order to do it we follow Ref.~[9] to perturb a spatial part of a metric
$$
ds^2=R^2(\tau)\[-d\tau^2+\(\d_{ij}+2\t E{}^T_{ij}\)dx^i\,dx^j\], \eq237
$$
$\tau$ is a conformal time.

If we expand Einstein equations for \rf237 \ up to linear terms, we get
$$
\frac{d^2E^T_{ij\vec k}}{d\tau^2}+2RH_K\frac{dE^T_{ij\vec k}}{d\tau}
+k^2E^T_{ij\vec k}=0, \quad K=0,1, \eq238
$$
where $\t E{}^T_{ij}$ is a spatial perturbation of the metric, $R(\tau)$ is a
scale factor depending on a conformal time, $E^T_{ij\vec k}$ is a $\vec k$
Fourier component $\t E{}^T_{ij}$
$$
\al
&\t E{}^T_{ij}=\frac1{(2\pi)^{3/2}}\int E^T_{ij\vec k}e^{-i\vec k\vec r}\,d^3\vec
k\\
&\vec k=(k_1,k_2,k_3), \quad |\vec k|^2=k^2, \quad \vec r=(x,y,z).
\eal
\eq239
$$
The important quantity is an amplitude of \gr\ wave corresponding to
$E^T_{ij}$, i.e.
$$
h_{ij\vec k}=RE^T_{ij\vec k}\,. \eq240
$$
$H_K$ is a Hubble \ct. We take $K=1$. The relation between a conformal time $\tau$ and
ordinary time~$t$ is given by
$$
\ealn{
R(\tau)&=-\frac1{H_1\tau}, \quad -\infty<\tau<0, &\rf241 \cr
R(t)&=R_1e^{H_1t}. &\rf242
}
$$
For $h_{ij\vec k}$ one gets
$$
\frac{d^2h_{ij\vec k}}{d\tau^2}-\frac2{\tau^2}\,\frac{dh_{ij\vec k}}{d\tau}
+k^2h_{ij\vec k}=0. \eq243
$$
This equation can be easily solved. For further investigations we need only
$$
\ealn{
h_{\vec k}&=h^i_{i\vec k} &\rf244 \cr
h_{\vec k}&=A_{\vec k}\(\tau|k| \cos(\tau|k|)-\sin(\tau|k|)\)+
B_{\vec k}\(\tau|k| \sin(\tau|k|)+\cos(\tau|k|)\) &\rf245
}
$$
and especially
$$
\frac{dh_{\vec k}}{d\tau}=A_{\vec k}\tau\sin(\tau|k|)+B_{\vec k}\tau
\cos(\tau|k|) \eq246
$$
where $A_{\vec k}$ and $B_{\vec k}$ are \ct s.

We need also a \q\ field time \dc ce (e.g.\ in a slow roll \ap). We use our
solution from the last point of Ref.~[2] (see Eq.~(14.416)), making some simplifications and
changing $t$ into~$\tau$:
$$
\P=\ln\(\sqrt{\frac{n+2}n}\sqrt{\frac\b{|\g|}}\)+\ln\(\frac{1-C(-\tau)^{\t k}}
{1+C(-\tau)^{\u\k}}\). \eq247
$$
Moreover what we really need is a derivative of $\P$.
$$
\P'=\frac{d\P}{d\tau}=\frac{2C{\t\k}(-\tau)^{{\u\k}-1}}{1-C^2(-\tau)^{2{\u\k}}} \eq248
$$
where
$$
\ealn{
{\t\k}&=\frac{\o B\sqrt{2n}}{H_1} &\rf248a \cr
C&=e^{\o B\sqrt{2n}t_0}\(R_1H_1\)^{\o B\sqrt{2n}/H_1}=
e^{\o B\sqrt{2n}t_0}\(R_1H_1\)^{\u\k}. &\rf248b
}
$$

Now we proceed to a \q\ fluctuation equation (see Ref.\ [7]):
$$
\d \ddot Q_{\vec k}+3H_1\d \dot Q_{\vec k}
+\(c^2_sk^2+R^2U''(Q)\)\d Q_{\vec k}= \frac1{2\o\b} h'_k\cdot \P'. \eq249
$$
One gets
$$
\al
\d \ddot Q_{\vec k}&+3H_1\d \dot Q_{\vec k}
+\(c^2_sk^2+R^2U''(Q)\)\d Q_{\vec k}\cr
&= \frac{C\t\k}{\o\b}\(A_{\vec k}\frac{(-\tau)^{{\u\k}-1}\sin(\tau|k|)}{1-C^2(-\tau)^{2{\u\k}}}
+B_{\vec k}\frac{(-\tau)^{{\u\k}-1}\cos(\tau|k|)}{1-C^2(-\tau)^{2{\u\k}}}\)\cr
&=\frac{C\t\k}{\o\b}\, C_{\vec k}\,\frac{\sin(\tau|k|+\d_{\vec
k})}{1-C^2(-\tau)^{2{\u\k}}}\,(-\tau)^{2{\u\k}}.
\eal
\eq250
$$
$\o\b$ is a normalization \ct\ in a definition of $Q=\frac\P{\o\b}$, 
$\frac1{\o\b{}^2}=\frac{8\pi |\o M|}{\mpl^2}$ (see \rf28 ). 

Let us consider $\d Q_{\vec k}$ as a Fourier component of a $q_0$ field
subject to the following initial conditions:
$$
\frac{dq_{0\vec k}(0)}{d\tau}=0=q_{0\vec k}(0). \eq251
$$
In this case one gets
$$
\frac{d^2q_{0\vec k}}{d\tau^2}+3H_1\frac{dq_{0\vec k}}{d\tau}
+\(c^2_sk^2+\frac{m_0^2}{H_1^2\tau^2}\)q_{0\vec k}=
\frac{(-\tau)^{{\u\k}-1}C{\t\k}}{\o\b\(1{-}C^2(-\tau)^{2{\u\k}}\)}C_{\vec k}\sin\(\tau|k|
{+}\d_{\vec k}\), \eq252
$$
$c_s$ is  a speed of sound in a \q, $C_{\vec k}$ and $\d_{\vec k}$ are \ct s.
Let us notice that $\tau=0$ corresponds to $t\to\infty$. Thus in some sense
we have asymptotic conditions for \gr\ waves. 

For further convenience it is good to change $\tau$ into $-\tau$. In this way
we get
$$
\frac{d^2q_{0\vec k}}{d\tau^2}-3H_1\frac{dq_{0\vec k}}{d\tau}
+\(c^2_sk^2+\frac{m_0^2}{H_1^2\tau^2}\)q_{0\vec k}=
-\frac{C{\t\k}\tau^{{\u\k}-1}}{\o\b\(1{-}C^2\tau^{2{\u\k}}\)}C_{\vec k}\sin\(\tau|k|
{-}\d_{\vec k}\), \eq253
$$
$$
\tau\in(0,+\infty), \quad \tau=\frac1{R_1H_1}e^{-H_1t}. \eq254
$$

The fluctuations (the field $q_0$) of a \q\ $Q$ are driven by \gr\ waves and
a \q\ field~$Q$. It is easy to see that we are interested in solutions for
small~$\tau$ (large~$t$). In this case one gets
$$
\frac{d^2q_{0\vec k}}{d\tau^2}-3H_1\frac{dq_{0\vec k}}{d\tau}
+\frac{m_0^2}{H_1^2\tau^2}q_{0\vec k}=-\frac{C{\t\k}\tau^{{\u\k}-1}}{\o\b}\,
C_{\vec k}\sin(\tau|k|-\d_{\vec k}). \eq255
$$
Let us calculate a \ct\ ${\t\k}$. One gets
$$
\al
{\t\k}&=\frac{\o B\sqrt{2n}}{H_1}=\(\frac n{n+2}\)^{n/4}\(\frac{|\g|}\b\)^{n/4}
\frac{\sqrt{\o M}}{\mpl}\cdot4(n+2)|\g|^{1/2}\\
&=8\sqrt{3\pi n}(n+2)\sqrt{\o M}\(\frac n{n+2}\)^{n/4}
\(\frac{m_{\u A}}{\mpl}\)^{(n+2)/2}\,\frac{|\uP|^{1/2}}{\a_s^{(n+1)}}
\(\frac{|\uP|}{\RG}\)^{n/2}.
\eal
\eq256
$$
If we use our simplified model with $n=14$, $\a_s^2=\a\dr{em}$, $m_{\u A}=
m\dr{EW}$, we get
$$
{\t\k}\cong10^{92}. \eq257
$$
Thus this is really a large number.

Moreover for small $\tau$ we have
$$
q_{o\vec k}(\tau)\simeq0,\quad \frac{dq_{0\vec k}}{d\tau}\simeq0. \eq258
$$
In this way we arrive to an equation
$$
\frac{d^2q_{0\vec k}}{d\tau^2}=-\frac{C\t\k}{\o\b}\(B_{\vec k}k\tau^{\u\k}+A_{\vec k}\tau
^{{\u\k}-1}\) \eq259
$$
and finally
$$
q_{0\vec k}(\tau)=-\frac{B_{\vec k}|k|C{\t\k}}{\o\b({\t\k}+2)({\t\k}+1)}\,\tau^{{\u\k}+2}
-\frac{A_{\vec k}C{\t\k}}{\o\b({\t\k}+1){\t\k}}\,\tau^{{\u\k}+1} \eq260
$$
or (taking under consideration that ${\t\k}$ is very large)
$$
\ealn{
q_{0\vec k}(\tau)&=-\frac{C}{\o\b\t\k} \(B_{\vec k}k\tau^{\u\k}+A_{\vec k}\tau^{{\u\k}-1}\)
& \rf261 \cr
q_{0\vec k}(\tau)&=-\frac{C}{\o\b\t\k} \biggl(A_{\vec k}(R_1H_1)^{-\o B\sqrt{2n}/H_1}
e^{-\o B\sqrt{2n}t}\cr
&\qquad\qquad{}+kB_{\vec k}(R_1H_1)^{-((\o B\sqrt{2n}/H_1)-1)}
e^{-\o B\sqrt{2n}t}\cdot e^{H_1t}\biggr) &\rf262 \cr
q_0(t,\vec r)&=\int q_{0\vec k}(t)e^{i\vec k\vec
r}\,d\vec k. &\rf263
}
$$
One gets
$$
\al
q_0(t,\vec r)&=-\frac{C}{\o\b\t\k} \biggl(g(\vec r)\(\frac{H_1}{R_1}\)^{\o B\sqrt{2n}/H_1}
e^{-\o B\sqrt{2n}t}\\
&\qquad\qquad{}+f(\vec r)\(\frac{H_1}{R_1}\)^{(\o B\sqrt{2n}/H_1)-1}
e^{-2\o B\sqrt{2n}t}\cdot e^{H_1t}\biggr) 
\eal \eq264
$$
where
$$
\ealn{
f(\vec r)&=\int A_{\vec k}e^{i\vec k\vec r}\,d^3\vec k
&\rf265 \cr
g(\vec r)&=\int |k|B_{\vec k}e^{i\vec k\vec r}\,d^3\vec k.
&\rf266
}
$$

$g(\vec r)$ and $f(\vec r)$ characterize a spatial \dc ce of \gr\ waves
background. For large $t$ we have to do with $g(\vec r)$, i.e.
$$
q_0(t,\vec r)=-\frac{C}{\o\b\t\k} g(\vec r)(R_1H_1)^{-((\o B\sqrt{2n}/H_1)-1)}
\cdot e^{-\o B\sqrt{2n}t}e^{H_1t}. \eq267
$$
In this way one gets from an energy-momentum tensor for $q_0$
$$
T_{\mu\nu}=\pa_\mu q_0\cdot \pa_\nu q_0-\tfrac12\eta_{\mu\nu}
\(\pa^\a q_0\cdot\pa_\a q_0+m_0^2q_0^2\) \eq268
$$
$$
\al
\rho_{q_0}&=T_{44}=\dot q^2_0-\frac12\(\dot q^2_0-
|\stpr\nabla q_0|^2-m_0^2q_0^2\)\cr
&=\frac12\,q_0^2\(\(H_1-\o B\sqrt{2n}\)^2+
|\stpr\nabla\ln g(r)|^2+m_0^2\)\cr
&=\frac{\rho\dr{gr}2n\o B{}^2 R_1^2\(\(H_1-\o B\sqrt{2n}\)^2+
|\stpr\nabla\ln g(r)|^2+m_0^2\)}{\o\b{}^2 e^{2\o B\sqrt{2n}t}}\,
e^{2H_1t}e^{\o B\sqrt{2n}t_0}
\eal
\eq269
$$
where
$$
\rho\dr{gr}=\tfrac12 g^2 \eq270
$$
is an energy density of primordial \gr\ waves.

For ${\t\k}=\frac{\o B\sqrt{2n}}{H_1}$ is a large number,
$$
\o B\sqrt{2n} \gg H_1 \eq271
$$
and $\rho_{q_0}$ is going to zero if $t\to\infty$. This density will be
frozen (no \dc ce on time) at an end of the second de Sitter phase $t=
t\dr{end}\gi{II}$, $t_0=t\dr{initial}\gi{II}$. In this way one gets
$$
\frac{\rho_{q_0}}{\rho\dr{gr}}=
\frac{2n\o B{}^2 R_1^2\(\(H_1-\o B\sqrt{2n}\)^2+
|\stpr\nabla\ln g(r)|^2+m_0^2\)}{\o\b{}^2\exp\(\frac{2\o B\sqrt{2n}N_1}{H_1}\)}\,
e^{2N_1} \eq272
$$
where $N_1$ is an amount of \il\ during the second de Sitter phase.
If $g(\vec r)=\text{const.}$, $\rho_{q_0}$ is isotropic and homogeneous.

The function $g(\vec r)$ can be written in a form
$$
g(\vec r)=g_0+\d g(\vec r) \eq273
$$
where $\d g(\vec r)$ is a small deviation and $g_0=\text{const.}$ One gets
$$
\stpr\nabla\(\ln g(\vec r)\)\simeq \frac1{g_0}\stpr\nabla\d g(\vec r). \eq274
$$
In this way one writes
$$
\rho_{q_0}=\rho_{q_0}^0+\d \rho_{q_0} \eq275
$$
where
$$
\al
\rho_{q_0}^0&=\frac{2n\o B{}^2 R_1^2\rho\dr{gr}\(\(H_1-\o B\sqrt{2n}\)^2+
m_0^2\)}{\o\b{}^2\exp\(\frac{2\o B\sqrt{2n}N_1}{H_1}\)}\,
e^{2N_1}\\
\d\rho_{q_0}&=|\stpr\nabla g(\vec r)|^2 \,
\frac{2n\o B{}^2 R_1^2\rho\dr{gr}}{\o\b{}^2\exp\(\frac{2\o B\sqrt{2n}N_1}{H_1}\)}\,
e^{2N_1}
\eal
\eq276
$$

According to our ideas the energy $\rho_{q_0}$ should be stored as a gas of
\q\ \pc s. The measurement of a low frequency speed of sound and a high
frequency speed of sound can give us $\eta$ and its spatial variation. In
this way we get also a square length of a gradient of $\d g(\vec r)$. If we
suppose that $\rho\dr{gr}$ is stored as a gas of gravitons (massless \pc s
with two polarization states) then one can write
$$
p\dr{gr}=\rho\dr{gr} \(\frac{T\dr{gr}}{\,\o T_0\,}\)^4 \eq277
$$
(see Eq.\ \rf231 ).

Probably we should also consider a gas of skewons. However, this is a
different story. Taking into account \rf235  \ and \rf236 \ we should expect
$\d \rho_{q_0}$ as extremely small and we can neglect it in our theory. 
Moreover, $\rho\dr{gr}$ is very important. Let us consider Eq.~\rf253 \ for
these values of $\tau$ for which the variable~$t$ is close to
$t\dr{end}\gi{I}$ (the end of the first de Sitter phase or beginning of the
second de Sitter phase). In this way one writes
$$
\tau=\frac1{H_1R_1}\,e^{-H_1t\dr{end}\gi I}+\xi=\tau_0+\xi,
\quad 0<\xi\ll 1. \eq278
$$
Taking into account that $\xi$ is very small we arrive to the following equation
$$
\frac{d^2q_{0\vec k}}{d\xi^2}  - 3H_1\,\frac{dq_{0\vec k}}{d\xi}
+\(c_s^2k^2+\frac{m_0^2}{H_1^2\tau_0^2}\)q_{0\vec k}
=-\frac{C_{\vec k}\tau_0}{2\xi\o\b}\sin\(\xi|k|+\o\d_{\vec k}\) \eq279
$$
where
$$
\o\d_{\vec k}=\tau_0|k|-\d_{\vec k}\,. \eq280
$$
We have also initial conditions
$$
\frac{dq_{0\vec k}}{d\xi}(0)=q_{0\vec k}(0)=0. \eq281
$$
Eq.\ \rf279 \ can be solved by a Laplace \tr\ method in the case of
$\o\d_{\vec k}=0$. 

Let
$$
c_s^2k^2+\frac{m_0^2}{H_1^2\tau_0^2}=b^2. \eq282
$$
One gets
$$
s^2\o q_{0\vec k}(s)-3H_1s\o q_{0\vec k}(s)+b^2\o q_{0\vec k}(s)=
-\frac{C_{\vec k}\tau_0}{2\o\b} \arctg\(\frac{|k|}s\) \eq283
$$
$$
\o q_{0\vec k}(s)=-\frac{C_{\vec k}\tau_0}{2\o\b}\,
\frac{\arctg\bigl(\frac{|k|}s\bigr)}{s^2-3H_1s+b^2} \eq284
$$
and
$$
q_{0\vec k}(\xi)=-\frac{C_{\vec k}\tau_0}{2\o\b}\intop_0^\infty
\frac{\arctg\bigl(\frac{|k|}s\bigr)e^{s\xi}}{s^2-3H_1s+b^2}\,ds. \eq285
$$
However, we cannot get any compact form of this solution. Moreover for we are
looking for a solution around zero, for $\o\d_{\vec k}=0$ we get
$$
-\frac{C_{\vec k}\tau_0}{2\xi}\sin(\xi|k|)\simeq
-\frac{C_{\vec k}\tau_0|k|}{2}\,. \eq286
$$

In this case the Laplace \tr\ method is successful:
$$
\ealn{
q_{0\vec k}(\xi)&=-\frac{C_{\vec k}\tau_0|k|}{2\o\b}\intop_0^\infty
\frac{e^{s\xi}}{s(s^2-3H_1s+b^2)}\,ds &\rf287 \cr
q_{0\vec k}(\xi)&=-\frac{C_{\vec k}\tau_0|k|}{2\o\b}
\(\frac1{b^2}+\frac{e^{s_1\xi}}{s_1(s_1-s_2)}+\frac{e^{s_2\xi}}{s_2(s_2-s_1)}
\) &\rf287a
}
$$
where $s_1$ and $s_2$ are roots of the polynomial
$$
s^2-3H_1s+b^2=0. \eq288
$$
Moreover $\xi$ is small. Thus we write
$$
e^{s_i\xi}\cong 1+s_i\xi, \quad i=1,2, \eq289
$$
and get
$$
q_{0\vec k}(\xi)=-\frac{C_{\vec k}\tau_0|k|}{2b^2\o\b}\(1+\frac{3H_1}{\sqrt\D}\)
\eq290
$$
if
$$
\D=9H_1^2-4b^2=9H_1^2-4\Bigl(c_s^2k^2+\frac{m_0^2}{H_1^2\tau_0^2}\Bigr)>0. \eq291
$$

If $\D<0$, one gets
$$
q_{0\vec k}(\xi)=-\frac{C_{\vec k}\tau_0|k|}{2b^2\o\b}\(1+\exp(\tfrac32 H_1\xi)
\(\frac{3H_1}{\om}\sin(\om\xi)-\cos(\om\xi)\)\) \eq292
$$
where
$$
4\om^2=-\D=4\Bigl(c_s^2k^2+\frac{m_0^2}{H_1^2\tau_0^2}\Bigr)-9H_1^2>0. \eq293
$$
For small $\xi$ one gets
$$
q_{0\vec k}(\xi)=-\frac{3 C_{\vec k}\tau_0|k|H_1}{4b^2\o\b}\,\xi, \eq294
$$
the \dc ce on $t$ is as follows:
$$
\xi=\frac{-1}{R_1H_1}\bigl(e^{-H_1t}-e^{-H_1t\dr{end}\gi{I}}\bigr). \eq295
$$
Moreover $t$ is close to \smash{$t\dr{end}\gi{I}$} and we get
$$
\xi\simeq \frac{e^{-H_1t\dr{end}\gi{I}}}{R_1}(t-t\dr{end}\gi{I}) \eq296
$$
for $t$ very close to \smash{$t\dr{end}\gi{I}$}.

In this way we get fluctuations of a \q\ in the moment of the first order
phase transition of the first de Sitter phase to the second de Sitter phase. 
 
Finally one gets
$$
\al
q_{0\vec k}(t)&=-\frac
{C_{\vec k}|k|e^{-H_1t\gi{I}\dr{end}}}
{2H_1R_1\(c_s^2|k|^2+R_1^2m_0^2e^{2H_1t\gi{I}\dr{end}}\)\o\b}\cr
&\times\(1+\frac{3H_1}{\sqrt{9H_1^2-4\bigl(c_s^2|k|^2+R_1^2m_0^2e^{2H_1t\gi{I}\dr{end}}\bigr)}}\)
\eal
\eq297
$$
for
$$
|k|<\frac1{2c_s}\sqrt{9H_1^2-4R_1^2e^{2H_1t\gi{I}\dr{end}}m_0^2}=k_0 \eq298
$$
with a condition
$$
3H_1>2R_1e^{H_1t\gi{I}\dr{end}}m_0 \eq299
$$
and
$$
q_{0\vec k}(t)=-\frac
{3 C_{\vec k}e^{-2H_1t\gi{I}\dr{end}}(t-t\gi{I}\dr{end})}
{4R_1^2\(c_s^2k^2+R_1^2m_0^2e^{2H_1t\gi{I}\dr{end}}\)\o\b}\, .
\eq300
$$

One finds
$$
\al
q_0(\vec r,t)&=-\frac{e^{-H_1t\gi{I}\dr{end}}}{2\o\b R_1}
\( \frac1{H_1}\intop_{|k|<k_0}
d^3\vec k\, e^{i\vec k\vec r}\frac
{C_{\vec k}|k|}{\(c_s^2k^2+R_1^2m_0^2e^{2H_1t\gi{I}\dr{end}}\)}\right.\cr
&\quad{}\times\(1+\frac{3H_1}{\sqrt{9H_1^2-4\(c_s^2k^2+R_1^2m_0^2e^{2H_1t\gi{I}\dr{end}}\)}}\)
\\
&\left.+\frac{3(t-t\gi{I}\dr{end})}{2R_1}\,e^{-H_1t\gi{I}\dr{end}}
\intop_{|k|>k_0}d^3\vec k\, e^{i\vec k\vec r}\frac
{C_{\vec k}}{\(c_s^2k^2+R_1^2m_0^2e^{2H_1t\gi{I}\dr{end}}\)}\)
\eal
\eq301
$$

Equation \rf301 \ gives us space-time fluctuations of a \q\ field after a
phase transition from the first to the second de Sitter phase. Let us notice
that \rf297 \ is singular for $k\to k_0$.

In an approach developed here we consider Eq.~\rf250 \ with zero initial
conditions in various \ap s. Moreover there is a different approach. In this
approach we write a solution to Eq.~\rf250
$$
q_{0\vec k}(t)=\t q_{0\vec k}(t)+\o q_{0\vec k}(t) \eq302
$$
where $\t q_{0\vec k}(t)$ is a solution of a homogeneous equation (a general
integral) and $\o q_{0\vec k}(t)$ is a special solution to inhomogeneous
equation. This different approach consists in taking $\t q_{0\vec
k}(t)\equiv0$ and considering only $\o q_{0\vec k}(t)$. In this case we get
nonzero initial conditions for \q\ fluctuations. However, this approach is not
unambigous. Our approach is unambigous. The zero initial conditions are in
some sense distinguished.

Let us calculate an energy density of this scalar field. In order to do this
let us write $q_0(\vec r,t)$ in a form
$$
q_0(\vec r,t)=A(\vec r)+B(\vec r)(t\gi{I}\dr{end}-t) \eq303
$$
where
$$
\ealn{
A(\vec r)&=-\frac{e^{-H_1t\gi{I}\dr{end}}}{2\o\b R_1H_1}
\intop_{|k|<k_0}
d^3\vec k\, e^{i\vec k\vec r}\frac
{C_{\vec k}|k|}{\(c_s^2k^2+R_1^2m_0^2e^{2H_1t\gi{I}\dr{end}}\)}\cr
&\quad{}\times\(1+\frac{3H_1}{\sqrt{9H_1^2-4\(c_s^2k^2+R_1^2m_0^2e^{2H_1t\gi{I}\dr{end}}\)}}\)
&\rf304 \cr
B(\vec r)&=-\frac{3 e^{-H_1t\gi{I}\dr{end}}}{4\o\b R_1^2}
\intop_{|k|>k_0}d^3\vec k\, e^{i\vec k\vec r}\frac
{C_{\vec k}}{\(c_s^2k^2+R_1^2m_0^2e^{2H_1t\gi{I}\dr{end}}\)}\,.
&\rf305
}
$$
One gets from the Eq.\ \rf268
$$
\rho_{q_0}=T_{44}=\frac12\(C(\vec r)+D(\vec r)(t\gi{I}\dr{end}-t)
+E(\vec r)(t\gi{I}\dr{end}-t)^2\) \eq306
$$
where
$$
\ealn{
C(\vec r)&=B^2(\vec r)+m_0^2A^2(\vec r)+\bigl|\stpr\nabla(A(\vec r))\bigr|^2 
&\rf307 \cr
D(\vec r)&=2\(\stpr\nabla (A(\vec r))\cdot \stpr\nabla (B(\vec r))
+m_0^2A(\vec r)B(\vec r)\) &\rf308 \cr
E(\vec r)&=\(\bigl|\stpr\nabla(B(\vec r))\bigr|^2+m_0^2B^2(\vec r)\).
&\rf309
}
$$
All these formulae are satisfied for $t$ close to $t\gi{I}\dr{end}$. In
general $\rho_{q_0}$ is anisotropic and inhomogeneous. For
$t\gi{I}\dr{end}-t$ is very small we get
$$
\rho_{q_0}=\frac12\(C(\vec r)+D(\vec r)(t\gi{I}\dr{end}-t)\). \eq310
$$

It is interesting to calculate a time $t(\vec r)$ for which $\rho_{q_0}$ is
zero ($t(\vec r)$ depends on a space point). One gets (neglecting gradients
of $A(\vec r)$ and $B(\vec r)$)
$$
t(\vec r)=t\gi{I}\dr{end}+\frac1{2m_0^2}\(\o f(\vec r)+
\frac{m_0^2}{\o f(\vec r)}\) \eq311
$$
where
$$
\o f(\vec r)=\frac{3H_1}{2R_1}\cdot
\frac{\intop_{|k|>k_0}d^3\vec k\, e^{i\vec k\vec r}\frac
{C_{\vec k}}{(c_s^2k^2+R_1^2m_0^2e^{2H_1t\gi{I}\dr{end}})}}
{\intop_{|k|<k_0}
d^3\vec k\, e^{i\vec k\vec r}\frac
{C_{\vec k}|k|}{(c_s^2k^2+R_1^2m_0^2e^{2H_1t\gi{I}\dr{end}})}
\biggl(1+\frac{3H_1}{\sqrt{9H_1^2-4(c_s^2k^2+R_1^2m_0^2e^{2H_1t\gi{I}\dr{end}})}}\biggr)}\,.
\eq312
$$

The conclusion from these calculations is that after a very short time an
energy density of \q\ fluctuations is going to zero (almost zero) and
probably is frozen on a very low level. A time of this process depends on a
space point. Moreover, after sufficiently long time (which is really very
short) we get a homogeneity and isotropy, as far as our rough calculations
advise us. Let us come back to Eq.~\rf253 . Our solutions obtained here are
for large $t$ and $t$ close to $t\gi{I}\dr{end}$. One can try to match them.
It seems reasonable to suppose that $\rho^0_{q_0}$ (see Eq.~\rf276 ) is equal
to this very low level energy density, mentioned above.

The \q\ field $q_0(\vec r,t)$ considered above can be a source of
fluctuations of an \ef\ \gr\ \ct\ $G\dr{eff}$.
$$
G\dr{eff}=G_N \exp\(-(n+2)\o\b q_0(\vec r,t)\) \eq313
$$
(see Eq.\ \rf28 ). If we use Eq.\ \rf267 , we get
$$
G\dr{eff}=G_N \exp\Biggl(-\frac
{\sqrt2\,(n+2)\sqrt{\rho_{q_0}}\,\o\b}
{\sqrt{\bigl((H_1-\o B\sqrt{2n})^2+\bigl|\stpr\nabla \ln g(\vec r)\bigr|^2
+m_0^2\bigr)}}\Biggr) \eq314
$$
and we connect contemporary fluctuations of an \ef\ \gr\ \ct\ to an energy
density of \q\ fluctuations. Taking a \q\ field from Eq.~\rf303 \ we get
$$
G\dr{eff}=G_N \exp\(-(n+2)\o\b\(A(\vec r)+B(\vec r)(t\gi{I}\dr{end}-t)\)\).
\eq315
$$
In this way an \ef\ \gr\ \ct\ depends on a speed of a sound in a \q.

We can consider some applications of a \q\ in astrophysics. First of all we
can develop a formalism concerning a macroscopic flow of a \q\ gas (i.e., a
relativistic hydrodynamics of a \q). For a \q\ field has an influence on an
\ef\ \gr\ \ct, such a flow can disturb \gr\ interactions between galaxies or
even stars. In this approach we should add some assumptions, e.g.\ that a \q\
energy density is an energy density of \q\ field in statistical field theory.
However, this is beyond a scope of this work.

Secondly, it is interesting to consider a Bose-Einstein condensation of \q\
\pc s. They are massive scalar \pc s. Under some assumptions (beyond this
theory) we can find a wave function of a condensat. This wave function can be
considered a field of a \q\ and afterwards enters the formula for
$G\dr{eff}$. In this way Einstein-Bose condensation can influence effective
\gr\ interactions between galaxies and stars. This is also beyond a scope of
the work.

As a typical astrophysical application we can consider a problem of
$N$~bodies with the \q\ field. In this problem $N$~points interact \gr ly
with an \ef\ \gr\ \ct\ depending on a distance. The simplest case is of
course a two bodies problem. However, this problem cannot be solved
analytically. It can be considered numerically using some codes for the
$N$~bodies problem. This is also beyond a scope of our work. Moreover, some
numerical solutions can be applied for galaxies movement in a cluster of
galaxies. 

In all of those problems we should consider \q\ fluctuations developed here
as a source of \q\ gas and Bose-Einstein condensat. In the third problem we
should also consider an interaction of a \q\ field with mass points.

It is interesting to consider a continuous distribution of a matter (a dust)
under \gr\ interaction and a \q. In this way we consider hydrodynamic
equation coupled to a \q\ field in a Newtonian physics limit (i.e.\
selfgravitating system with $G\dr{eff}$ depending on a \q).

\section{Conclusions}
In the paper we consider a mass of a \q\ \pc, a speed of sound in a \q\ and
several solutions to \q\ equations coming to the interesting behaviour of an
\ef\ \gr\ \ct. Some of issues of our theory are considered in
self-interacting Brans-Dike theory [8].

The fraction $\eta$ can be connected with the part of an energy density of
\q\ field $Q$ in such a way that it is a fraction of fluctuation energy density
around an equilibrium~$Q_0$. In this way the $q_0$ field which can evolve due
to an evolution of~$Q$ and due to fluctuations of primordial \gr\ waves will
be a source of a gas of thermalized \pc s. It seems that an approach with a
boson equation of state is more appropriate to consider. 

\def\ii#1 {\item{[#1]}}
\section{References}
\setbox0=\hbox{[11]\enspace}
\parindent\wd0

\ii1 {Kalinowski} M. W.,
{\it A short note on a total amount of inflation\/},
arXiv: hep-th/0307299v2.

\ii2 {Kalinowski} M. W., 
{\it Nonsymmetric Fields Theory and its Applications\/},
World Scientific, Singapore, New Jersey, London, Hong Kong 1990.

\item{} {Kalinowski} M. W., 
{\it Nonsymmetric Kaluza--Klein (Jordan--Thiry) Theory in the electromagnetic
case}\/,
Int. Journal of Theor. Phys. {\bf31}. p.~611 (1992).

\item{} {Kalinowski} M. W., 
{\it Nonsymmetric Kaluza--Klein (Jordan--Thiry) Theory in a general
nonabelian case}\/,
Int. Journal of Theor. Phys. {\bf30}, p.~281 (1991).

\item{} {Kalinowski} M. W., 
{\it Can we get confinement from extra dimensions}\/,
in: Physics of Elementary Interactions (ed. Z.~Ajduk, S.~Pokorski,
A.~K.~Wr\'oblewski), World Scientific, Singapore, New
Jersey, London, Hong Kong 1991.

\item{} {Kalinowski} M. W., 
{\it Scalar fields in the nonsymmetric Kaluza--Klein (Jordan--Thiry)
theory}, arXiv: hep-th/0307242.

\ii3  {Kalinowski} M. W., 
{\it A warp factor in the nonsymmetric Kaluza--Klein (Jordan--Thiry)
Theory\/}, 
arXiv: hep-th/0306157v2.

\ii4 {Kalinowski} M. W., 
{\it Dynamics of Higgs' field and a quintessence
in the nonsymmetric Kaluza--Klein (Jordan--Thiry) Theory}, 
arXiv: hep-th/0306241v1. 

\ii5 {Emden} R.,
{\it Gaskugeln: Anwendungen der mechanischer W\"arme\/},
B.~G.~Teubner, Leipzig und Berlin 1907.

\item{} {Chandrasekhar} S.,
{\it Introduction to the Study of Stellar Structure\/},
New York, Dover 1939.

\ii6 {Lemke} H.,
{\it \"Uber die Differentialgleichungen, welche den Gleichgewichtszustand
eines gasf\"ormigen Himmelsk\"orpers bestimmen, dessen Teile gegeneinander
nach dem Newtonschen Gesetze gravitieren\/},
Journal f\"ur Mathematik, Bd.~142, Heft~2, S.~118 (1913).

\ii7 {Steinhardt} P. J.,
{\it The quintessential introduction to dark energy\/},
Lecture given at {\sl The search of dark matter and dark energy in the
Universe\/}, The Royal Society, January 22--23, 2003.

\item{} {DeDeo} S., {Caldwell} R. R., {Steinhardt} P. J.,
{\it Effects of the sound speed of \q\ on the microwave background and large
scale structure\/},
arXiv: astro-ph/0301284v2.

\item{} {Erickson} J. K., {Caldwell} R. R., {Steinhardt} P. J.,
{Armendriz-Picon} C., {Mukha\-nov}~V., 
{\it Measuring the speed of sound of \q\/},
arXiv: astro-ph/0112438v1.

\item{} {Rahul} D., {Caldwell} R. R., {Steinhardt} P. J.,
{\it Sensitivity of the cosmic microwave background anisotropy to initial
conditions in \q\ cosmology\/},
arXiv:\break astro-ph/0206372v2.

\item{} {Caldwell} R. R., {Steinhardt} P. J.,
{\it The imprint of \gr\ waves in models dominated by a dynamical cosmic
scalar field\/},
arXiv: astro-ph/9710062v1.

\item{} {Steinhardt} P. J.,
{\it Quintessential cosmology and cosmic acceleration\/},
see Steinhardt's home WWW page.

\ii8 {Chakraborty} S., {Chakraborty} N. C., {Debnath} U.,
{\it A \q\ problem in self-interacting Brans-Dicke theory\/},
arXiv: gr-gc/0305103v1.

\item{} {Chakraborty} S., {Chakraborty} N. C., {Debnath} U.,
{\it Quintessence problem and\break Brans-Dicke theory\/},
arXiv: gr-qc/0306040v1.

\ii9 {Liddle} A. R., {Lyth} D. H., 
{\it Cosmological Inflation and Large-Scale Structure}\/,
Cambridge Univ. Press, Cambridge 2000.

\end